\RequirePackage{fix-cm}
\documentclass[smallcondensed,natbib,final]{svjour3}     
\smartqed  
\usepackage{graphicx}
\usepackage{booktabs}
\usepackage{color}
\usepackage[citecolor=blue,urlcolor=blue,breaklinks]{hyperref}
\usepackage{amssymb}
\usepackage{amsmath}
\usepackage{epsfig}
\usepackage{float}
\usepackage{rotating}
\usepackage{psfrag}
\usepackage{subfigure}
\usepackage{wrapfig}
\usepackage{placeins}
\usepackage{comment}

\newcommand{\eg}{{e.g.,}}
\newcommand{\alf}{{Alfv{\'e}n }}

\newcommand{\sA}{ {\scriptscriptstyle{\rm A}} }
\newcommand{\sSun}{ {\scriptscriptstyle{\rm \odot}} }

\newcommand{\be}{\begin{equation}}
\newcommand{\ee}{\end{equation}}
\newcommand{\bea}{\begin{eqnarray}}
\newcommand{\eea}{\end{eqnarray}}
\newcommand{\bean}{\begin{eqnarray*}}
\newcommand{\eean}{\end{eqnarray*}}
\newcommand{\equationname}{Eq.~}

\journalname{Living Rev. Sol. Phys.}
\begin{document}

\title{Extended MHD modeling of the steady solar corona and the solar wind}

\author{Tamas I.\ Gombosi
\and 
Bart van der Holst
\and 
Ward B.\ Manchester
\and 
Igor V.\ Sokolov
}

\institute{T. I.\ Gombosi \at 
Center for Space Environment Modeling, \\
University of Michigan \\
2455 Hayward, Ann Arbor, MI 48109, USA \\
\email{tamas@umich.edu}
\and
B. van der Holst \at 
CSEM, University of Michigan \\
\email{bartvand@umich.edu}
\and
W. B.\ Manchester IV\at 
CSEM, University of Michigan \\
\email{chipm@umich.edu}
\and
I. V.\ Sokolov \at 
CSEM, University of Michigan \\
\email{igorsok@umich.edu}
}

\date{Received: date / Accepted: date}

\maketitle

\begin{abstract}
The history and present state of large-scale magnetohydrodynamic (MHD) modeling of the solar corona and the solar wind with steady or quasi-steady coronal physics is reviewed. We put the evolution of ideas leading to the recognition of the existence of an expanding solar atmosphere into historical context. The development and main features of the first generation of global corona and solar wind models are described in detail. This historical perspective is also applied to the present suite of global corona and solar wind models. We discuss the evolution of new ideas and their implementation into numerical simulation codes. We point out the scientific and computational challenges facing these models and discuss the ways various groups tried to overcome these challenges. Next, we discuss the latest, state-of-the art models and point to the expected next steps in modeling the corona and the interplanetary medium.
\end{abstract}


\newpage
\setcounter{tocdepth}{3}
\tableofcontents

\newpage

\section{Introduction}
\label{sec:intro}

It was realized thousands of years ago that the space between heavenly objects must be filled by something that is much lighter than the materials found on Earth. \textit{Aether} (ancient Greek for \textit{light}) was one of the primordial deities in Greek mythology. He was the personification of the upper air. Later the name ``aether'' was used by Greek philosophers to describe a very light fifth element. Plato mentioned that ``there is the most translucent kind which is called by the name of aether.'' Aristotle introduced a new ``first'' element to the system of the classical elements (Earth, Fire, Air, Water). He noted that the four terrestrial classical elements were subject to change and naturally moved linearly. The first element however, located in the celestial regions and heavenly bodies, moved circularly and had none of the qualities the terrestrial classical elements had. It was neither hot nor cold, neither wet nor dry. With this addition the system of elements was extended to five and later commentators started referring to the new first one as the fifth and also called it aether. Medieval scholastic philosophers granted aether changes of density, in which the bodies of the planets were considered to be more dense than the medium which filled the rest of the universe. Later, scientists speculated about the existence of aether to explain light and gravity. Isaac Newton also suggested the existence of an aether \citep{Newton:1718a}. 

The idea that the Sun might be the source of corpuscular radiation was first suggested by \cite{Carrington:1860a}. On September 1, 1859, \cite{Carrington:1860a} and \cite{Hodgson:1860a} independently observed a huge white-light solar flare. Less than a day later telegraph communications were severely disrupted during a planetary-scale magnetic storm. At the same time a great aurora was seen, even in Rome, a truly exceptional event \citep{Green:2006a}. On September 2, the telegraph line between Boston and Portland (Maine) operated on ``celestial power,'' without batteries \citep{Loomis:1860a}. \cite{Carrington:1860a} suspected a causal relationship between the solar flare, the magnetic storm, and the aurora, and he suggested a continuous stream of solar particles as a way to connect these phenomena.

Carrington's suspicion, however, was not universally shared. Thirty-three years after the so-called ``Carrington event'', Lord Kelvin (William Thomson), in a Presidential Address to the Royal Society, argued that the Sun was incapable of powering even a moderate-sized magnetic storm: His argument was based on his unwillingness to think outside the box and his not accepting the possibility that the Sun might be powered by a process that was not understandable in terms of ``classical'' physics. He was trying to explain the Sun's energy production within the framework of coal burning. His logic lead him to the conclusion that the Sun could not be older than a million years. Rather than questioning his own basic assumptions, he questioned Darwin's estimates that some fossils might be hundreds of millions of years old. At the end of his talk he confidently concluded, ``It seems as if we may also be forced to conclude that the supposed connection between magnetic storms and sunspots is unreal, and that the seeming agreement between the periods has been a mere coincidence'' \citep{Thomson:1893a}. This reminds us of the old quote from Michel de Montaigne: ``Nothing is so firmly believed as what we least know.''

The idea of a charged corpuscular radiation emanating from the Sun was next suggested by \cite{FitzGerald:1892a} who wrote, ``a sunspot is a source from which some emanation like a comet's tail is projected from the Sun ... . Is it possible, then, that matter starting from the Sun with the explosive velocities we know possible there, and subject to an acceleration of several times solar gravitation, could reach the Earth in a couple of days?''

A few years later, \cite{Roentgen:1896a} discovered that cathode rays could cause crystals to fluoresce, pass through solid objects, and affect photographic plates. A few years later, \cite{Lodge:1900a} suggested that magnetic storms were due to ``a torrent or flying cloud of charged atoms or ions''; that auroras were caused by ``the cathode ray constituents ... as they graze past the polar regions''; that comet tails could not be accounted for by solar electromagnetic radiation pressure but could be accounted for by particle radiation emanating from sunspots ``like a comet's tail'' and ``projected from the Sun'' with an ``average velocity [of] about 300 miles per second''; and finally, that ``there seems to be some evidence from auroras and magnetic storms that the Earth has a minute tail like that of a comet directed away from the Sun'' (quotes selected by \citealp{Dessler:1967a}).

The electron was discovered at the very end of the 19th century \citep{Thomson:1897a} and the concept of an electrically neutral solar radiation composed of oppositely charged particles was not introduced until the middle of the 1910s \citep{Birkeland:1916a}. He wrote: ``From a physical point of view it is most probable that solar rays are neither exclusively negative nor positive rays, but of both kinds.'' 

From his geomagnetic surveys, \cite{Birkeland:1908a} realized that auroral activity was nearly uninterrupted, and he concluded that the Earth was being continually bombarded by ``rays of electric corpuscles emitted by the Sun.'' His work was, however, pretty much ignored at the time. Some seventy years later, \cite{Dessler:1984a} wrote an intriguing article arguing that personality conflicts between Sidney Chapman and Scandinavian scientists, as well as British imperial arrogance, were partly responsible for ignoring the pioneering works of Birkeland, Enskog, St{\o}rmer and Alfv{\'e}n.

In the early 1930s, \cite{Chapman:1931a, Chapman:1931b, Chapman:1932a, Chapman:1932b, Chapman:1933a} published the first quantitative model of an infinitely conducting quasi-neutral plasma beam and its interaction with a magnetic dipole. Chapman was fundamentally a mathematician and he always tried to simplify physics phenomena to treatable mathematical problems. Ferraro was a PhD student looking for a thesis topic. \cite{Akasofu:1995a} quotes Chapman on why he did not consider a continuous plasma stream from the Sun: ``He [Lindeman] said it [a stream of gas] must consist of charges of opposite signs in practically equal numbers, so that it could hold together. Lindemann never tried to develop what would be the consequences on the Earth of the impact of such a stream of gas. I made an attempt at that while I was Professor at Manchester in 1919--1924, but unfortunately I started at the wrong end; I tried to find out what would be the steady state, as if the stream had been going on forever. It didn't work out; so I was still wanting to find out what would happen, and this was the subject I proposed for Ferraro.'' In plain English, Chapman did not consider the case of a continuous solar wind because he could not solve this problem mathematically.

\section{Early ideas}
\label{sec:early}

\subsection{The puzzle of comet tails}
\label{sec:comets}

\begin{figure}[htb]
\centering
\includegraphics[width=0.8\textwidth]{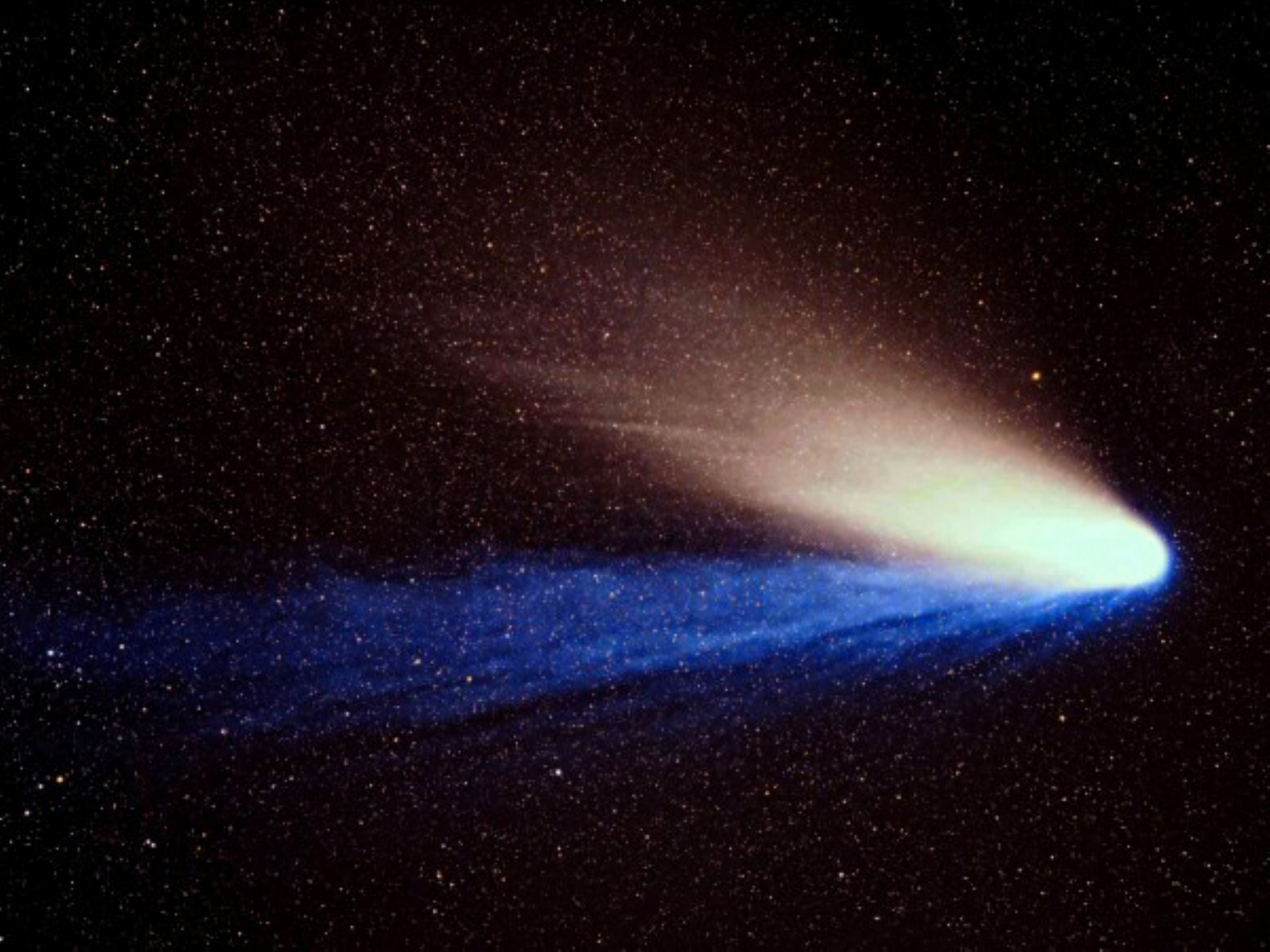}
\caption{Comet Hale-Bopp (1997) showing two distinct tails -- a broad dust tail (white) and a narrow ion tail (blue). (Source: \url{http://www.tivas.org.uk/solsys/tas_solsys_comet.html)}}
\label{fig:hale-bopp}
\end{figure}

The idea of a continuous corpuscular radiance emanating from the Sun did not resurface until the 1950s. Interestingly, it was the observation of comet tails that lead to the re-emergence of the idea. Comets exhibit two distinct types of tails. A nice example is comet Hale-Bopp, shown in \figurename~\ref{fig:hale-bopp}. 

By the 1950s, it had been recognized that the broad and curved dust tails (also called Type~II tails) usually lag behind the Sun-comet line (opposite to the direction of cometary orbital motion). Since the gravity of the cometary nucleus is negligible, the motion of the dust particles is controlled by an interplay between solar gravity and solar radiation pressure. Assuming that dust grains have a more or less constant mass density, $\rho_d$, the anti-sunward radiation pressure, $F_{rad}$ is inversely proportional to the square of the heliocentric distance, $d_h$,  and proportional to the cross section of the dust grain, $\pi a^2$ ($a$ is the equivalent radius of a spherical dust grain), $F_{rad} \propto a^2/d_h^2$. The sunward pointing gravitational force is proportional to the particle mass ($4\pi \rho_d a^3/3$) and inversely proportional to the heliocentric distance: $F_{grav} \propto a^3/d_h^2$. Since the two forces point in the opposite direction, have the same heliocentric dependence, but exhibit different dependence on the grain size ($F_{rad} \propto a^2$ and $F_{grav} \propto a^3$), the resulting effect is a complex ``dust mass spectrometer'' where each particle size is moving under the influence of its own ``reduced solar gravity.'' This effect results in the broad and curved dust tail.

For active comets the straight, narrow plasma tails (also called Type~I tails) are $10^7$--$10^8$ km long and, within a few degrees, always point away from the Sun. Observation of various tail structures as they moved down the Type~I tail determined that the acceleration in plasma tails ranged from about 30 to 300~cm s$^{-2}$ and occasionally even larger. This value was some three orders of magnitude larger than any acceleration due to solar radiation pressure. Some major process was missing. In order to account for the observed large acceleration \cite{Biermann:1951a} postulated the existence of a continuous ``solar corpuscular radiation'' composed of electrons and ions. Assuming that the antisunward acceleration of small irregularities in Type~I comet tails was due to Coulomb collisions between electrons in a radially outward plasma flow from the Sun and newly ionized cometary particles, \cite{Biermann:1951a} inferred a solar wind density and velocity of $n_{sw} \approx 1,000$ cm$^{-3}$ and $u_{sw} \approx 1,000$ km/s that represents a particle flux $\sim$500 times larger than was later observed.

The important consequence of Biermann's (\citeyear{Biermann:1951a}) idea was that if solar corpuscular radiation is responsible for the antisolar acceleration of comet tails then the Sun evidently emits solar corpuscular radiation in all directions at all times. This follows from the fact that  comets ``fill'' the heliosphere; antisunward pointing comet tails were observed over the poles of the Sun as well as at low heliographic latitudes. In addition, comets come by as frequently at sunspot minima as at sunspot maxima. Yet none fail to show an antisolar Type I tail. This means that interplanetary space must be completely filled with solar corpuscular radiation.

Bierman's idea was pretty much ignored by the solar physics community. The fact that the Sun has a million-degree corona was first discovered by \cite{Grotian:1939a} and \cite{Edlen:1941a} by identifying the coronal lines as transitions from low-lying metastable levels of the ground configuration of highly ionized metals (the green FeXIV line at 530.3 nm, but also the red line FeX at 637.4 nm). In the mid-1950s \cite{Chapman:1957a} calculated the properties of a gas at such a temperature and determined it was such a superb conductor of heat that it must extend way out into space, beyond the orbit of Earth. However, Chapman -- and others -- considered a static corona that was gravitationally bound by the Sun. If the Earth is moving at about 30 km/s along its orbit around the Sun, the interaction between the Earth and the stationary solar corona could only result in very minor geomagnetic disturbances. Large geomagnetic storms, however, were already well observed at that time.

\subsection{Solar wind}
\label{sec:wind}

Sometime in 1957, Ludwig Biermann visited John Simpson's group at the University of Chicago. While in Chicago he had extensive discussions with one of Simpson's postdocs who was working on the problem of cosmic ray modulation in the solar system. Biermann explained his comet tail idea to the postdoc, Eugene Parker. According his own recollection, Parker started to think about the Biermann--Chapman puzzle: how to reconcile Chapman's (\citeyear{Chapman:1957a}) hot, highly conducting corona with Biermann's (\citeyear{Biermann:1951a}) idea of a continuously outward streaming fast solar corpuscular radiation. Parker's solution to the Biermann--Chapman puzzle was the idea of the continuous expansion of the hot solar corona, the solar wind \citep{Parker:1958a}.

\begin{figure}[htb]
\centering
\includegraphics[width=0.8\textwidth]{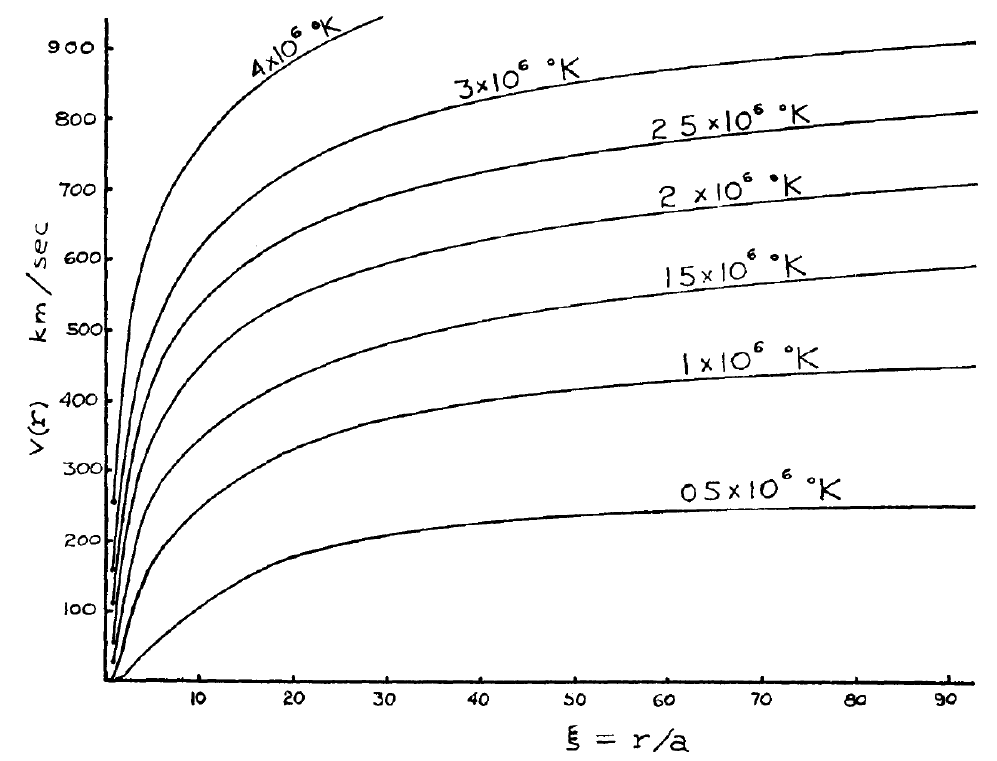}
\caption{Spherically symmetric hydrodynamic expansion velocity $v(r)$ of an isothermal solar corona with temperature $T_0$ plotted as a function of $r/a$, where $a$ is the radius of the base of the solar corona and has been taken to be $10^6$ km \citep{Parker:1958a}.}
\label{fig:parker}
\end{figure}

In his seminal paper, \cite{Parker:1958a} pointed out that the static corona has a finite pressure at infinity that exceeds the pressure of the interstellar medium by a large factor. He concluded that the solar corona cannot be static (however, as it was later pointed out by \cite{Velli:1994a} and \cite{DelZanna:1998a}, the situation is quite complicated). Next, he considered a steady-state spherically symmetric isothermal corona that expands with a velocity $v(r)$, where $r$ is the heliocentric distance. In addition, he assumed that the electron and ion temperatures were identical and approximated the plasma pressure by $p=2nkT_0$, where $n(r)$ is the ion number density, $T_0$ is the ion temperature, and $k$ is the Boltzmann constant. Using the conservation of particle flux and the momentum equation, he obtained the following analytic relation for the plasma flow velocity:
\begin{equation}
  \left[\frac{v^2}{v_m^2} - \ln\left(\frac{v^2}{v_m^2}\right)\right]
    = 4\ln\left(\frac{r}{a}\right) 
    + \left(\frac{v_{esc}^2}{v_m^2}\right)\left(\frac{a}{r}\right)
    - 4 \ln\left(\frac{v_{esc}^2}{v_m^2}\right)
    - 3 + \ln 256 \,,
\label{eq:parker.1}
\end{equation}
where $a$ is the radius of the hot coronal base (Parker used a value of $a=10^6$ km), $v_{esc}$ is the escape velocity at radius $a$ ($v_{esc}^2=2GM_\sSun/a$, where $G$ is the gravitational constant and $M_\sSun$ is the solar mass), while $v_m$ is the most probable velocity ($v_m^2=2kT_0/m_p$, where $m_p$ is the proton mass). In \equationname(\ref{eq:parker.1}), the integration constant was chosen to ensure that the velocity is real and positive for all $r > a$ values. In addition, only solutions with $v_0^2 \ll {2kT_0}/{m}$ and $v(r\rightarrow\infty)>v_m$ solutions were considered, since the expansion velocity at the base of the hot corona, $v_0$ was assumed to be very small and Parker was interested in a hydrodynamic escape solution. This solution was later called the ``solar wind.'' \figurename~(\ref{fig:parker}) shows the class of escaping corona solutions \citep{Parker:1958a}.

It is interesting to note that in Parker's (\citeyear{Parker:1958a}) solution the expansion velocity reaches the most probable particle velocity at the heliocentric distance of $r_c=a v_{esc}^2/4 v_m^2$. Since it is assumed that the outflow velocity is small at the coronal base, this means that hydrodynamic outflow can only take place if the $v_m^2>v_{esc}^2/4$ condition is met. In plane English, one needs a very hot corona for the solar wind to exist.

In his original paper \cite{Parker:1958a} also considered the effect of the general outflow of solar gas upon the solar dipole magnetic field. He assumed that there are no field-free regions in the Sun, so that each volume element of gas flowing outward from the Sun will carry the embedded magnetic field lines with it. The field lines, being embedded in both the Sun and the ejected gas, will be stretched out radially as the gas moves away from the Sun. The radial configuration will be as universal as the outward-gas motion, which is responsible for it. \cite{Parker:1958a} suggested that the gas flowing out from the Sun is not field-free but carries with it magnetic lines of force originating in the Sun. Hence, he predicted a radial solar magnetic field, falling off approximately as $1/r^2$ in interplanetary space. The magnetic field lines will follow an Archimedean spiral described by
\begin{equation}
  \frac{r}{R_s} - \ln\left(\frac{r}{R_s}\right)
    = 1 + \frac{v_s}{R_s \Omega_\sSun} \left(\phi - \phi_s\right) \,,
\label{eq:parker.2}
\end{equation}
where $\Omega_\sSun$ is the angular velocity of the Sun, $\phi$ is the solar azimuth angle and $\phi_s$ is the azimuth angle of the magnetic field line at the distance $R_s$ (\cite{Parker:1958a} assumed a value of $R_s = 5 R_\sSun$). Using this simple model \cite{Parker:1958a} also expressed the magnetic field vector at an arbitrary point outside the $r=R_s$ sphere:
\begin{eqnarray}
  B_r \left(r, \theta, \phi\right) 
    &=& B\left(\theta, \phi_s\right) \left(\frac{R_s}{r}\right)^2 \nonumber \\
  B_\theta \left(r, \theta, \phi\right) &=& 0 \nonumber \\
  B_\phi \left(r, \theta, \phi\right) 
    &=& B\left(\theta, \phi_s\right) 
    \frac{\Omega_\sSun \left(r-R_s\right)\sin\theta}{v_s}
    \left(\frac{R_s}{r}\right)^2 \,,
\label{eq:parker.3}
\end{eqnarray}
where $v_s$ is the asymptotic speed of the solar wind that is reached at a heliocentric distance, where $B\left(\theta, \phi_s\right)$ is the radial component of the magnetic field at the origin of the field line (on the $r=R_s$ sphere). 

\begin{figure}[htb]
\centering
\includegraphics[width=0.6\textwidth]{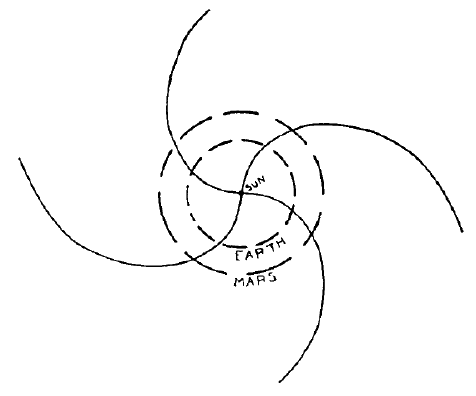}
\caption{Solar equatorial interplanetary magnetic field lines carried by outward-streaming plasma with velocity $10^3$ km/sec \citep{Parker:1958a}.}
\label{fig:imf}
\end{figure}

At large heliocentric distances (where $r\gg R_s$) the radial magnetic field component decreases as $B_r\propto1/r^2$, while the azimuthal component only decreases as $B_\phi\propto1/r$. This means that in the interplanetary space the magnetic field lines become more and more ``wound up'' as one moves further and further away from the Sun. This heliospheric magnetic field configuration is now referred to as the ``Parker spiral.'' A schematic of the interplanetary magnetic field line topology is shown in \figurename~\ref{fig:imf} \citep{Parker:1958a}.

Another important consequence of Parker's (\citeyear{Parker:1958a}) solution was that, due to the conservation of mass flux, it predicted that the particle density in the interplanetary medium would decrease as $n \propto 1/r^2$, much slower than the exponential decrease obtained in the case of a hydrostatic atmosphere. This result was consistent with Biermann's (\citeyear{Biermann:1951a}) conclusions, but it was very different than the accepted model predictions.

Opposition to Parker's (\citeyear{Parker:1958a}) hypothesis on the solar wind was immediate and strong. He submitted the manuscript to the \textit{Astrophysical Journal} in early January of 1958. Here is how Parker remembers the events \citep{Parker:2001a}: 

``... sometime in the late Spring of 1958 the referee's report on the original paper appeared, with the suggestion that the author familiarize himself with the subject before attempting to write a scientific article on solar corpuscular radiation. There was no specific objection to the arguments or calculations in the submitted paper. The author pointed out to the editor the absence of any substantive objections by the referee. The editor, Subrahmayan Chandrasekhar, sent the paper to a second referee. Sometime in the summer the second referee responded that the paper was misguided and recommended against publication, again with no specific criticism except that the author was obviously unfamiliar with the literature. Again, the author's response to the editor was to note that there was no substantive criticism, no specific fault pointed out. Some days later Chandrasekhar appeared in the author's office with the paper in his hand and said something along the lines of, `Now see here, Parker, do you really want to publish this paper?' I replied that I did. Whereupon he said, `I have sent it to two eminent experts in the field and they have both said that the paper is misguided and completely off the mark.' I replied that my problem with the referees was that they were clearly displeased, but had nothing more to say. Chandrasekhar was silent for a moment and than he said, `All right, I will publish it.' And that is how the paper without the words `solar wind' finally appeared in the November issue of the Astrophysical Journal.''

\begin{figure}[htb]
\centering
\includegraphics[width=0.8\textwidth]{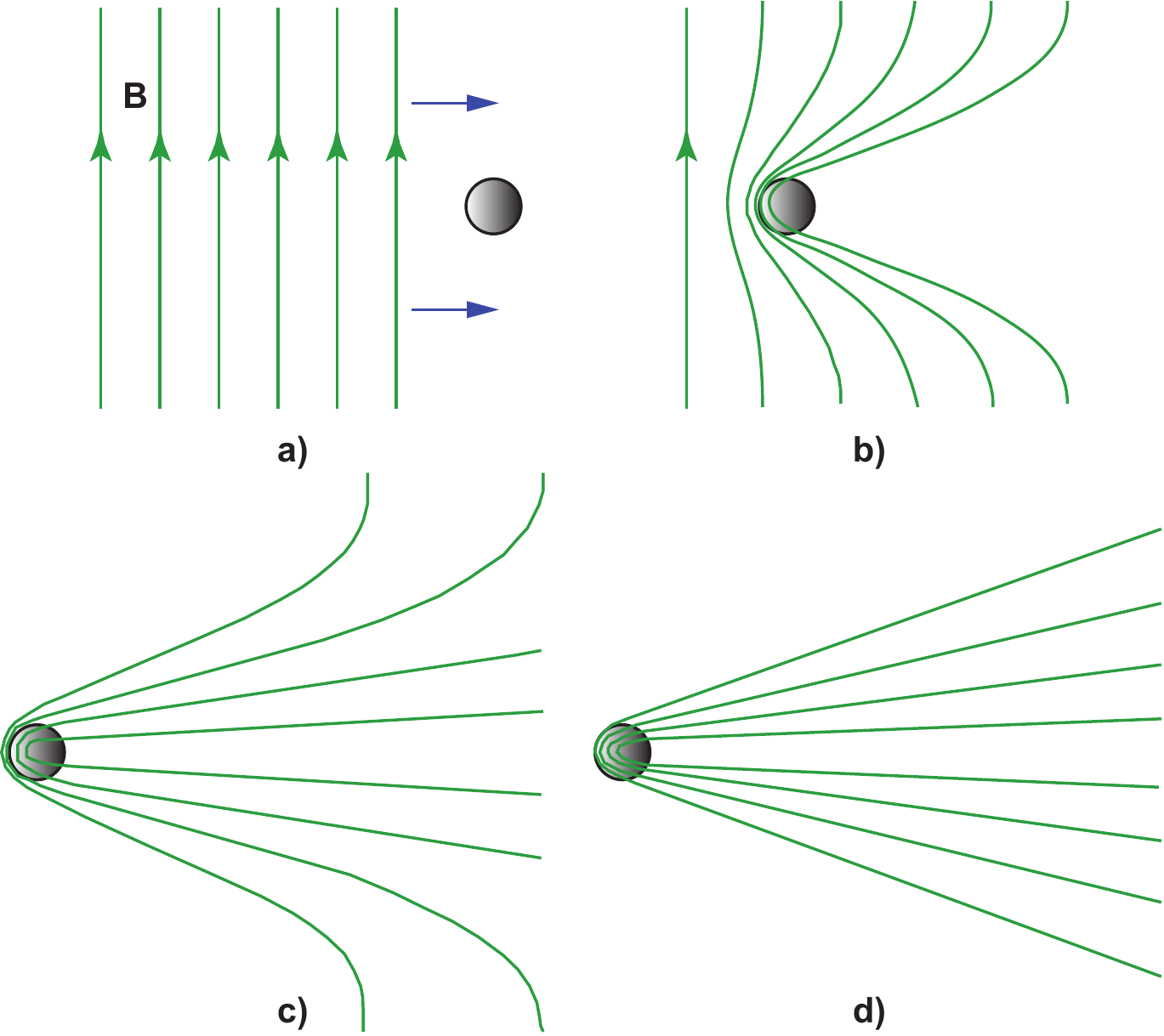}
\caption{Alfv{\'e}n's scenario of the formation of comet tails. A plasma beam with a frozen-in magnetic field approaches the head of a comet (panel a); the field is deformed (panels b and c); the final state is reached when the beam has passed (panel d) \citep{Alfven:1957a}.}
\label{fig:alfvenfig}
\end{figure}


Just about the same time when \cite{Parker:1958a} published his paper solving the Biermann-Chapman puzzle, \cite{Alfven:1957a} pointed out that the plasma densities inferred by \cite{Biermann:1951a} were inconsistent with observed coronal densities (assuming that the plasma density decreases with the square of the heliocentric distance as the solar corpuscular radiation moves outward). \cite{Alfven:1957a} offered an alternative explanation for the formation of cometary plasma tails that is depicted in \figurename~\ref{fig:alfvenfig}. The main assumption in this model was that the solar corpuscular radiation was carrying a ``frozen-in'' magnetic field that ``hangs up'' in the high density inner coma, where the solar particles strongly interact with the cometary atmosphere and consequently the solar plasma flow considerably slows down. This interaction results in a ``folding'' of the magnetic field around the cometary coma that creates the long plasma tail. Disturbances along the folded magnetic field lines propagate as magnetohydrodynamic waves and they can reach velocities of 100 km/s even if the solar plasma has a density of $\sim5$ cm$^{-3}$. It is interesting to note that \alf accepted Biermann's (\citeyear{Biermann:1951a}) idea of the continuous plasma outflow from the Sun and naturally concluded that this plasma outflow must carry an embedded magnetic field, but he did not worry about the origin of such a plasma flow.

\subsection{Solar breeze?}
\label{sec:chamberlain}

\begin{figure}[htb]
\centering
\includegraphics[width=0.8\textwidth]{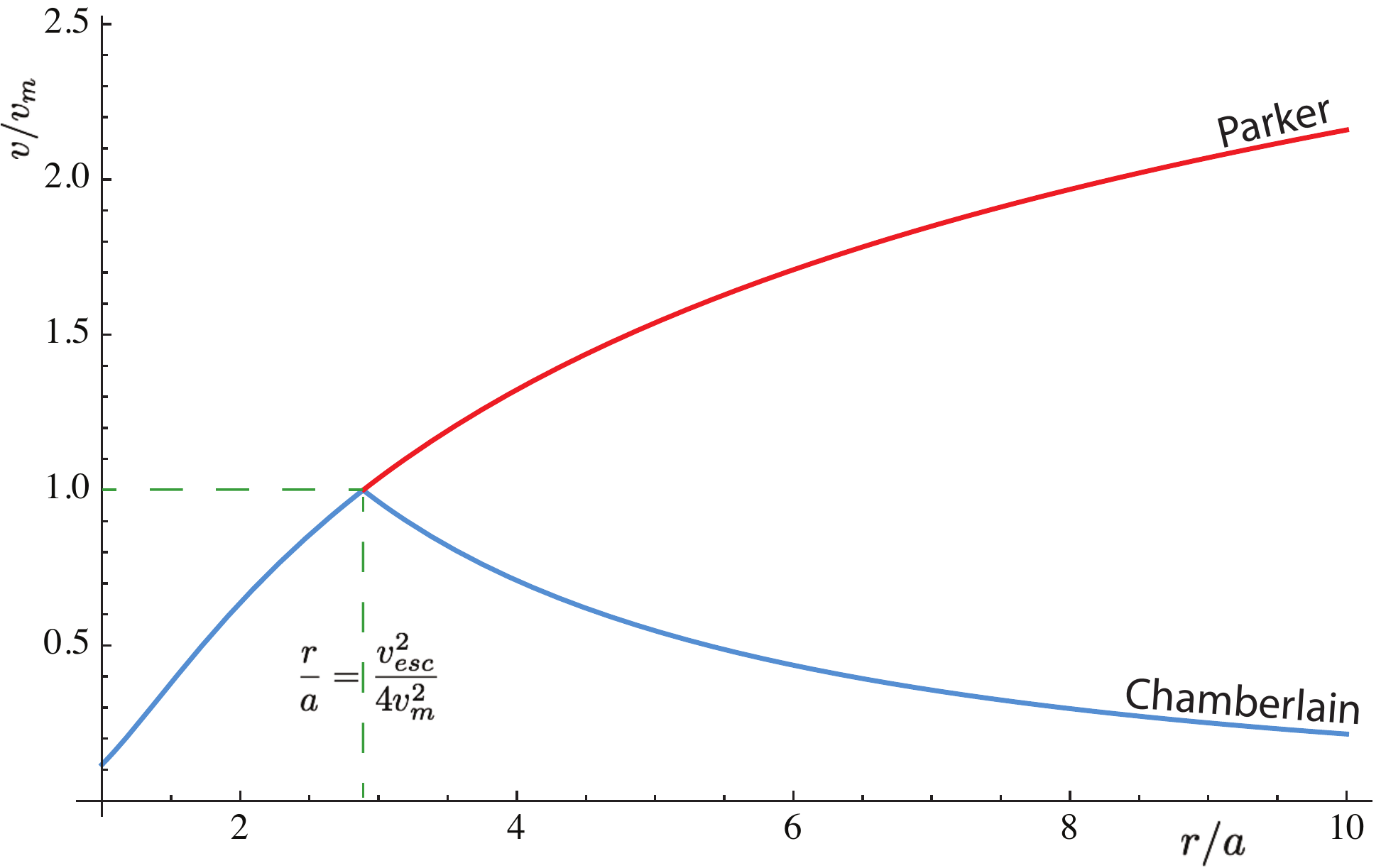}
\caption{Comparison of the \cite{Parker:1958a} and \cite{Chamberlain:1960a} solar corona solutions.}
\label{fig:wind-breeze}
\end{figure}

Parker's (\citeyear{Parker:1958a}) paper generated swift negative reaction. The community was not ready to give up the idea of a hydrostatic corona and accept a continuously escaping solar atmosphere. 

The most prominent critic of Parker was Joseph Chamberlain, who received his PhD from the University of Michigan in 1952. At that time he was on the faculty of the University of Chicago, the same institution where Chandrasekhar and Parker worked. In September 1959, less than a year after Parker's (\citeyear{Parker:1958a}) paper was published, he submitted a paper which strongly criticized Parker's (\citeyear{Parker:1958a}) solar wind solution. \cite{Chamberlain:1960a} pointed out that \equationname(\ref{eq:parker.1}) had two solutions that met the condition that the velocity was small at the coronal base. In addition to Parker's (\citeyear{Parker:1958a}) solution, a second solution described a solar corona that started to expand, but the expansion gradually stopped beyond the critical point ($r_c=a v_{esc}^2/4 v_m^2$). A comparison of the \cite{Parker:1958a} and \cite{Chamberlain:1960a} solutions is shown in \figurename~\ref{fig:wind-breeze}.

Chamberlain's (\citeyear{Chamberlain:1960a}) solution provided a hydrostatic coronal density distribution and the expansion velocity returned to zero at infinity. This solution ``saved'' most features of the prevailing hydrostatic atmosphere model and it was greeted with great relief by most of the solar community. The disagreement between Parker and Chamberlain became quite heated. In 1959 when the debate took place Parker was a junior (untenured) faculty member at the University of Chicago and, as a result of the ongoing controversy with a more senior faculty member, his tenure case became more challenging. The controversy was eventually resolved -- at least in the eyes of the space physics community -- by the beginning of the space age, when \textit{in-situ} observations proved the existence of the solar wind.

\cite{Gringauz:1960a} analyzed the results of the Lunik-2 spacecraft and determined an interplanetary corpuscular particle flux of about $2\times10^8$ ions/cm$^2$/s: ``Starting from 9.30 hr Moscow time on 13 September 1959 up to the moment of the container of the second space rocket reaching the moon the container was recorded as passing through a positive ion flux (in all probability protons) with energies exceeding 15 eV; $\Phi \sim 2 \times 10^8$ ions/cm$^2$/s.'' In a May 1962 presentation in Washington, \cite{Gringauz:1964a}  used observations of the first deep space probe (Venera-1) to estimate the speed of the solar corpuscular radiation to be about 400 km/s. In addition, measurements by \cite{Bonetti:1963a} on Explorer~10 confirmed these initial results. However, because the Lunik, Venera and Explorer observations were short-term, the doubters had some wiggle-room and the observations were not regarded as definitive. 

\begin{figure}[htb]
\centering
\includegraphics[width=0.9\textwidth]{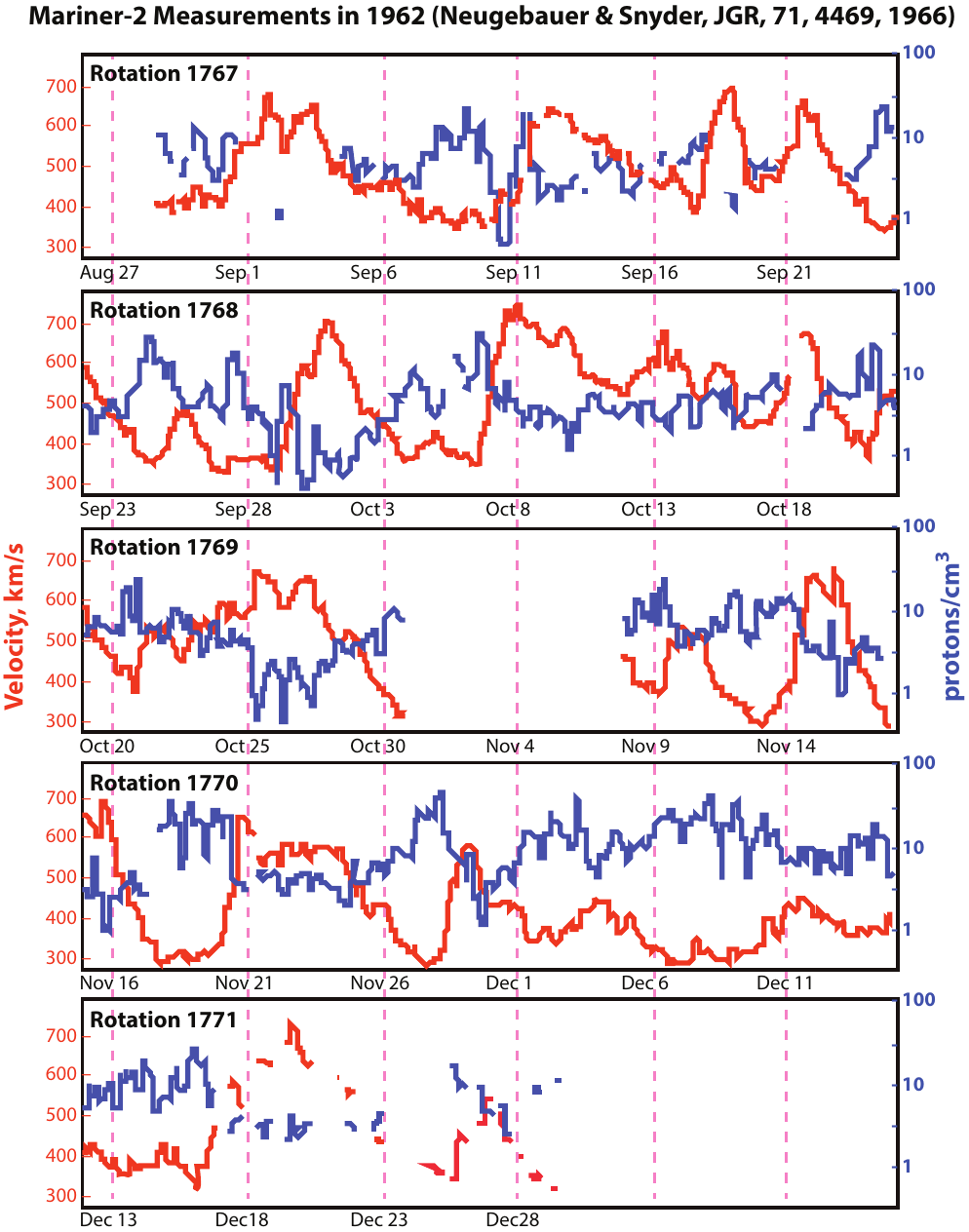}
\caption{Mariner-2 observations of the continuous solar wind \citep{Neugebauer:1966a}.
\label{fig:mariner}
}
\end{figure}

It was not until the end of 1962, when the first Mariner 2 results were reported \citep{Neugebauer:1966a}, that the existence of the solar wind was widely accepted. The observed solar wind had typical proton densities of $5$ to $20$ cm$^{-3}$ and velocities between 300 and 700 km/s (see \figurename~\ref{fig:mariner}). The observed temperature was in the range of $3\times10^4$--$3\times10^5$ K. These observations confirmed the predictions of the \cite{Biermann:1951a}--\cite{Alfven:1957a}--\cite{Parker:1958a} theory and proved that the concept of a static, slowly evaporating corona was incorrect \citep{Chapman:1957a, Chamberlain:1960a}.

It is interesting to note Chamberlain's reaction to the eventual acceptance of the solar wind concept. Even some thirty years later, he tried to show that those who advocated the solar wind concept were right for the wrong reasons and they advocated inappropriate solutions. Here is a quote from Chamberlain's \citeyear{Chamberlain:1995a} paper:

``In the early days of solar-wind theory, \cite{Parker:1958a} appeared to be influenced by two seminal hypotheses: (a) Biermann's conclusion that the behavior of comet tails was governed more by the interplanetary medium than by solar-radiation pressure, and (b) Chapman's advocacy of a static solar corona that was heated to great distances by conduction. Parker showed that a static corona was untenable and then constructed a \textit{primitive} hydrodynamic model, which he labeled the \textit{solar wind}, that would account for Biermann's analysis of comet tails. I describe this solar-wind model as `primitive' because its temperature distribution, instead of being derived physically, was characterized by a polytropic index, which simplified the model enormously.

At this point I was skeptical that Parker's supersonic solutions were \textit{realistic} \citep{Chamberlain:1960a} and, to investigate the problem, I wrote down the three equations -- now called the \textit{solar-wind equations} -- for a hydrodynamic solar corona that was heated from below by thermal conduction \citep{Chamberlain:1961a}. I restricted my investigation to slow (subsonic) expansion, or models in which the parameter $\varepsilon$ (described below) was 0. ln a jocular spirit, I referred to those solutions as \textit{solar-breeze models}, and I suggested that they were merely the hydrodynamic counterpart to Waterston-Jeans evaporative escape.

Shortly afterward, a solar-wind model, based on the same three equations but encompassing supersonic expansion, was developed by Scarf and Noble. (The pair of papers, \cite{Noble:1963a} and \cite{Scarf:1965a}, are hereinafter abbreviated by `SN-I' and `SN-II,' respectively.) But their model, although quite acceptable as an illustrative case, is not physically accurate, even within the constraints of its own simplifying assumptions.''

\section{First numerical models of the solar corona}
\label{sec:first_corona}

\subsection{Numerical solution}
\label{sec:numerical}

\begin{figure}[htb]
\centering
\includegraphics[width=0.6\textwidth]{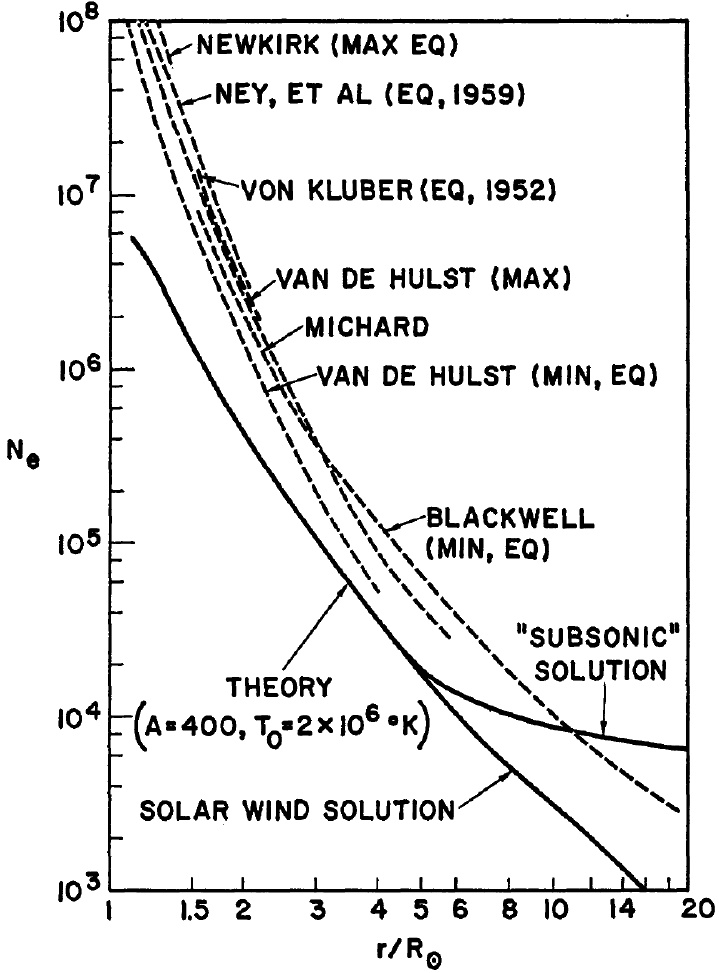}
\caption{Comparison between simulated and observed electron densities in the inner corona (neglecting heat conduction and viscous effects) \citep{Scarf:1965a}.
\label{fig:scarf}
}
\end{figure}

The first numerical solution of the solar wind equations was carried out by Scarf and Noble \citep{Noble:1963a, Scarf:1965a}, who solved the spherically symmetric Navier-Stokes equations including heat conduction and viscosity effects. Instead of considering a polytropic corona, they solved the energy equation, thus making the problem more challenging. The set of differential equations had a singularity when the local solar wind speed, $v(r)$, reached the value of $\sqrt{k T(r)/m_p}$. In order to avoid numerical problems, \cite{Noble:1963a} integrated the equations from the Earth inward. This was made possible by the fact that at that time the solar wind conditions were observed. \cite{Noble:1963a} chose 1~AU values of $n=3.4$ cm$^{-3}$, $v=352$ km/s, and $T=2.77\times10^5$ K.

A typical solution without heat conduction and viscous effects is shown in \figurename~\ref{fig:scarf}. We note that even in this relatively simple case the transonic solar wind solution describes the observed electron density profile within a factor of 2 or 3, a surprisingly good agreement. \cite{Scarf:1965a} also considered the ``subsonic solution,'' representing the solar breeze \citep{Chamberlain:1960a}. When they took into consideration heat conduction and viscosity, \cite{Scarf:1965a} were able to get excellent agreement with observations (however, in effect, they increased the number of free parameters, so the improved agreement is not really surprising). A similar solution with heat conduction but without viscosity was also obtained by \cite{Whang:1965a}.

\subsection{Two-fluid model}
\label{sec:2fluid}

The first two-fluid (separate electron and proton equations) was published by \cite{Sturrock:1966a}. They realized that as the solar wind leaves the vicinity of the Sun collisional coupling between electrons and ions becomes weak and the electron and ion temperatures ($T_e$ and $T_i$) can deviate from each other. They developed a two-temperature model where the plasma remains quasi-neutral ($n_e\approx n_i$) and current-free (${\bf u}_e={\bf u}_i$), but the two temperatures can be different. \cite{Sturrock:1966a} still considered a spherically symmetric problem and neglected magnetic field effects, but their model represented a step forward. By combining the continuity, momentum, and energy equations they derived separate ``heat equations'' for electrons and ions \citep{Sturrock:1966a}:
\begin{equation}
  \frac{3}{2}\frac{1}{T_s}\frac{dT_s}{dr} - \frac{1}{n}\frac{dn}{dr}
    = \frac{1}{\Phi k T_s} \frac{d}{dr}\left(r^2\kappa_s\frac{dT_s}{dr}\right)
    + \frac{3}{2}\frac{\nu_{ei}}{v}\frac{T_t - T_s}{T_s} \,,
\label{eq:sturrock}
\end{equation}
where $s=e,i$ refers to either electrons or ions, the $t$ subscript refers to the other species, $\Phi=nvr^2$ is the spherical particle flux, $v$ is the plasma flow speed, $\kappa_s$ is the heat conductivity of the appropriate species and $\nu_{ei}$ is the electron-ion energy transfer collision frequency. \cite{Sturrock:1966a} used the \cite{Chapman:1954a}  approximation for the collision frequency ($\nu_{ei} = 8.5\times 10^{-2} nT_e^{-3/2}$, where $n$ is given in units of cm$^{-3}$) and the \cite{Spitzer:1962a} conductivities ($\kappa_e = 6\times10^{-7} T_e^{5/2}$ s$^{-1}$ and $\kappa_i = 1.4\times10^{-8} T_i^{5/2}$ s$^{-1}$).

\begin{figure}[htb]
\centering
\includegraphics[width=0.45\textwidth]{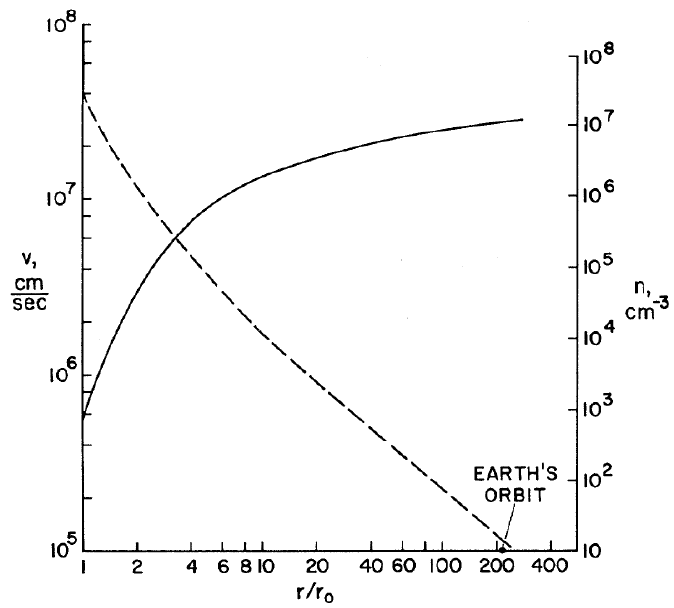}
\includegraphics[width=0.5\textwidth]{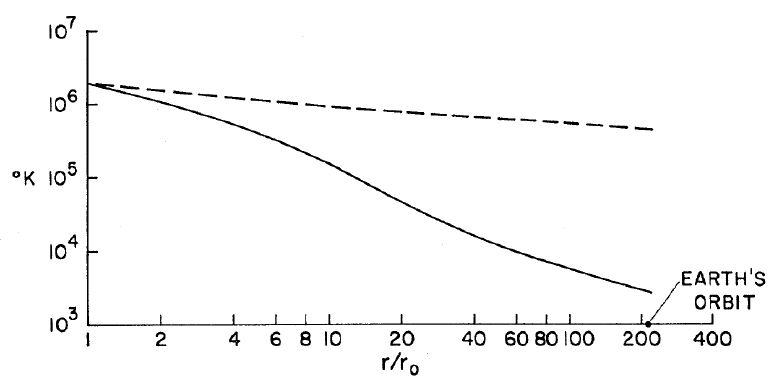}
\caption{Left panel: Flow velocity (solid line) and electron density (broken line) as a function of radial distance (in units of solar radii) from the center of the Sun. Right panel: Electron (broken line) and ion (solid line) temperatures as a function
of radial distance \citep{Sturrock:1966a}.
\label{fig:sturrock}
}
\end{figure}

Figure~\ref{fig:sturrock} shows the two-temperature solar wind solution \citep{Sturrock:1966a}. At Earth orbit they obtained $v=270$ km/s, $n=13$ cm$^{-3}$, $T_i=2800$ K and $T_e=4.6\times10^5$ K. While these values were in the right ballpark, the wind speed was too slow, the density was too high, and the two temperatures were quite a bit off ($T_e$ too high and $T_i$ too low). The authors lamented that ``We are left with the problem of understanding why the solar-wind parameters do not always agree with our model.'' In other words, it was Nature's fault that it did not agree with the predictions of the model...

\subsection{Potential magnetic field}
\label{sec:potential}

By the late 1960s, it has become clear that the magnetic field plays a major role in determining the density as well as the velocity and temperature structure of the corona. The first models used the observed line-of-sight component of the photospheric magnetic field to determine the magnetic field of the solar corona in the current-free (or potential-field) approximation \citep{Schatten:1968a, Schatten:1969a, Schatten:1969b, Altschuler:1969a, Newkirk:1970a, Schatten:1971a}.

The potential field model is based on the fundamental assumption that the magnetic field above the photosphere is current free ($\nabla\times{\bf B}=0$ when $r>R_\sSun$) and therefore the coronal magnetic field can be represented by a magnetic scalar potential, $\psi$:
\begin{equation}
  {\bf B} = -\nabla \psi \,.
\label{eq:pf1}
\end{equation}
Since there are no magnetic monopoles ($\nabla\cdot{\bf B}=0$) we obtain
\begin{equation}
  \nabla^2 \psi = 0 \,.
\label{eq:pf2}
\end{equation}

The solution of \equationname (\ref{eq:pf2}) can be written as an infinite series of spherical harmonics:
\begin{equation}
  \psi\left(r,\theta,\phi\right) = R_\sSun \sum_{n=1}^\infty \sum_{m=0}^n 
    \left(\frac{R_\sSun}{r}\right)^{n+1} \left[
    g_n^m \cos{m\phi} + h_n^m \sin{m\phi} \right] P_n^m\left(\theta\right) \,,
\label{eq:pf}
\end{equation}
where the coefficients $g_n^m$ and $h_n^m$ need to be determined from solar line-of-sight observations and $P_n^m\left(\theta\right)$ are the associated Legendre polynomials with Schmidt normalization:
\begin{equation}
  \frac{1}{4\pi} \int_0^\pi d\theta \int_0^{2\pi} d\phi P_n^m\left(\theta\right)
    P_{n^\prime}^{m^\prime}\left(\theta\right) 
    \left\{\begin{array}{c} \cos{m\phi} \\ \sin{m\phi}\end{array}\right\}
    \left\{\begin{array}{c} \cos{m^\prime\phi} \\ \sin{m^\prime\phi}\end{array}\right\}
    \sin \theta = \frac{1}{2n+1} \delta_{nn^\prime} \delta_{mm^\prime} \,.
\label{eq:legendre}
\end{equation}
With the help of the magnetic scalar potential, the magnetic field vector can be obtained anywhere above the photosphere:
\begin{equation}
  {\bf B} = \left(B_r,B_\theta,B_\phi\right) =
    \left(-\frac{\partial\psi}{\partial r}, -\frac{1}{r}\frac{\partial\psi}{\partial\phi},
    -\frac{1}{r \sin\theta}\frac{\partial\psi}{\partial\phi}\right) \,.
\label{eq:b-vector}
\end{equation}

It was recognized by \cite{Schatten:1969b} that the coronal magnetic field follows the current-free potential solution between the photosphere and a ``source surface'' (located at $r=R_s$) where the potential vanishes and the magnetic field becomes radial. This requires a network of thin current sheets for $r\ge R_s$ \cite[cf.\ ][]{Schatten:1971a}. Such a potential field can be described with the help of Legendre polynomials that define the magnetic potential between two spherical shells, located at $r=R_\sSun$ and $r=R_s$, each containing a distribution of magnetic sources \citep{Altschuler:1969a}:
\begin{equation}
  \psi_{s}\left(r,\theta,\phi\right) = 
    R_\sSun \sum_{n=1}^\infty f_n\left(r\right) \sum_{m=0}^n 
    \left[g_n^m \cos{m\phi} + h_n^m \sin{m\phi} \right] P_n^m\left(\theta\right) \,,
\label{eq:source-surface}
\end{equation}
where
\begin{equation}
  f_n\left(r\right) = 
    \frac{\left(\frac{R_s}{R_\sSun}\right)^{2n+1}}
    {\left(\frac{R_s}{R_\sSun}\right)^{2n+1}-1}
    \left(\frac{R_\sSun}{r}\right)^{n+1} 
    - \frac{1}{\left(\frac{R_s}{R_\sSun}\right)^{2n+1}-1}
    \left(\frac{r}{R_\sSun}\right)^{n} \,.
\label{eq:func1}
\end{equation}
For the optimal radius of the source surface \cite{Schatten:1969b} found $R_s=1.6\,R_\sSun$, while \cite{Altschuler:1969a} recommended $R_s=2.5\,R_\sSun$. Today, most potential field source surface (PFSS) models use the $R_s=2.5\,R_\sSun$ value. We note that at the source surface
\begin{equation}
  f_n\left(R_s\right) = 
    \frac{1}{\left(\frac{R_s}{R_\sSun}\right)^{2n+1}-1}
    \left(\frac{R_s}{R_\sSun}\right)^{n} 
    - \frac{1}{\left(\frac{R_s}{R_\sSun}\right)^{2n+1}-1}
    \left(\frac{R_s}{R_\sSun}\right)^{n} = 0 \,.
\label{eq:func2}
\end{equation}
With the help of expression~(\ref{eq:source-surface}), the magnetic field components can be written as
\begin{eqnarray}
  B_{r}\left(r,\theta,\phi\right) & = &
    \left(\frac{R_\sSun}{r}\right) \sum_{n=1}^\infty n \, f_n\left(r\right) \sum_{m=0}^n 
    \left[g_n^m \cos{m\phi} + h_n^m \sin{m\phi} \right] P_n^m\left(\theta\right)
\nonumber \\ && \hspace{-0.65in}
    + \sum_{n=1}^\infty 
    \frac{\left(\frac{R_s}{R_\sSun}\right)^{2n+1}}
    {\left(\frac{R_s}{R_\sSun}\right)^{2n+1}-1} 
    \left(\frac{R_\sSun}{r}\right)^{n+2}
    \sum_{m=0}^n \left[g_n^m \cos{m\phi} + h_n^m \sin{m\phi} \right] 
    P_n^m\left(\theta\right)
\label{eq:source-surface-field-r}
\end{eqnarray}
\begin{equation}
  B_{\theta}\left(r,\theta,\phi\right) = -
    \left(\frac{R_\sSun}{r}\right) \sum_{n=1}^\infty f_n\left(r\right) \sum_{m=0}^n 
    \left[g_n^m \cos{m\phi} + h_n^m \sin{m\phi} \right] 
    \frac{d P_n^m\left(\theta\right)}{d\theta}
\label{eq:source-surface-field-theta}
\end{equation}
\begin{equation}
  B_{\phi}\left(r,\theta,\phi\right) =
    \left(\frac{R_\sSun}{r\sin\theta}\right) 
    \sum_{n=1}^\infty f_n\left(r\right) \sum_{m=0}^n 
    m\,\left[g_n^m \sin{m\phi} - h_n^m \cos{m\phi} \right] 
    P_n^m\left(\theta\right)
\label{eq:source-surface-field-phi}
\end{equation}
At the source surface the magnetic field becomes radial because $f_n\left(R_s\right)=0$ and
\begin{equation}
  B_{r}\left(R_s,\theta,\phi\right) =
    \sum_{n=1}^\infty 
    \frac{\left(\frac{R_s}{R_\sSun}\right)^{n-1}}
    {\left(\frac{R_s}{R_\sSun}\right)^{2n+1}-1}
    \sum_{m=0}^n \left[g_n^m \cos{m\phi} + h_n^m \sin{m\phi} \right] 
    P_n^m\left(\theta\right) \,.
\label{eq:source-surface-field-r-Rs}
\end{equation}

\begin{figure}[htb]
\centering
\includegraphics[width=0.6\textwidth]{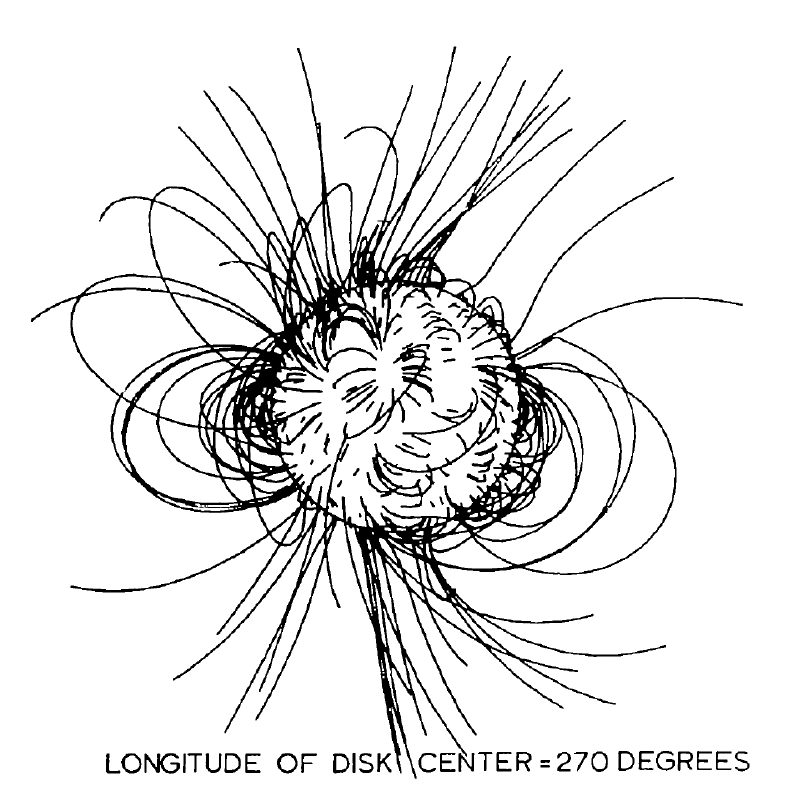}
\caption{Magnetic field line map obtained with the potential field source surface (PFSS) model \citep{Altschuler:1969a}.
\label{fig:altschuler}
}
\end{figure}

Outside the source surface it is assumed that the radial flow of the solar wind carries the magnetic field outward into the heliosphere. This region is not described by the PFSS model. Between the photosphere and the source surface the magnetic field vector components can be described by expressions (\ref{eq:source-surface-field-r}), (\ref{eq:source-surface-field-theta}) and (\ref{eq:source-surface-field-phi}). The inner boundary condition at the photosphere is obtained from the observed line-of-sight magnetic field components using a least square fit \cite[cf.\ ][]{Hoeksema:1982a}. These measurements are used to determine the expansion coefficients $g_n^m$ and $h_n^m$. The outer boundary condition at the source surface is that the field is normal to the source surface, consistent with the assumption that it is then carried outward by the solar wind. This condition is automatically satisfied by our selection of the $f_n$ function (\equationname(\ref{eq:func1})). An example of the PFSS solution is shown in \figurename~\ref{fig:altschuler}.

\subsection{Expanding magnetized corona}
\label{sec:expandingcorona}

\begin{figure}[htb]
\centering
\includegraphics[width=0.8\textwidth]{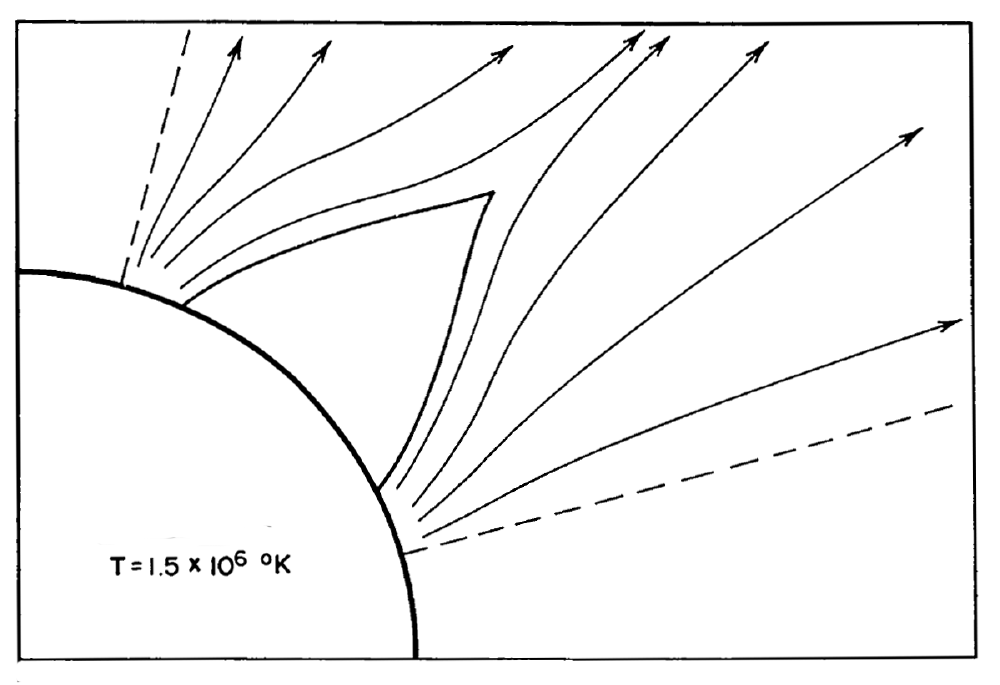}
\caption{Analytic solution of a helmet configuration \citep{Pneuman:1968a}.
\label{fig:pneuman_helmet}
}
\end{figure}

Nearly simultaneously with the development of the source surface models the effect of the coronal expansion on the magnetic field was also explored \citep{Pneuman:1966a, Pneuman:1968a, Pneuman:1969a, Pneuman:1971a}. In a series of papers \cite{Pneuman:1966a, Pneuman:1968a, Pneuman:1969a} analytically investigated how centrifugal, pressure gradient, and magnetic forces impact the flow of an infinitely conducting fluid where the magnetic field lines and plasma flow lines must coincide in the corotating frame.

An early example of a numerical model of the corona with solar wind is by \cite{Pneuman:1971a}.  As in the Parker solution \citep[]{Parker:1963a}, this model possesses the high temperature ($T > 1$ MK) and comparatively high density ($\rho \approx 10^{-16}$ gcm$^{-3}$) plasma whose pressure cannot be held in equilibrium by solar gravity or the pressure of the interstellar medium.  Consequently, the coronal plasma expands rapidly outward achieving supersonic speeds within a few solar radii, and in doing so forms the solar wind. 

\begin{figure}[htb]
\centering
\includegraphics[width=0.6\textwidth]{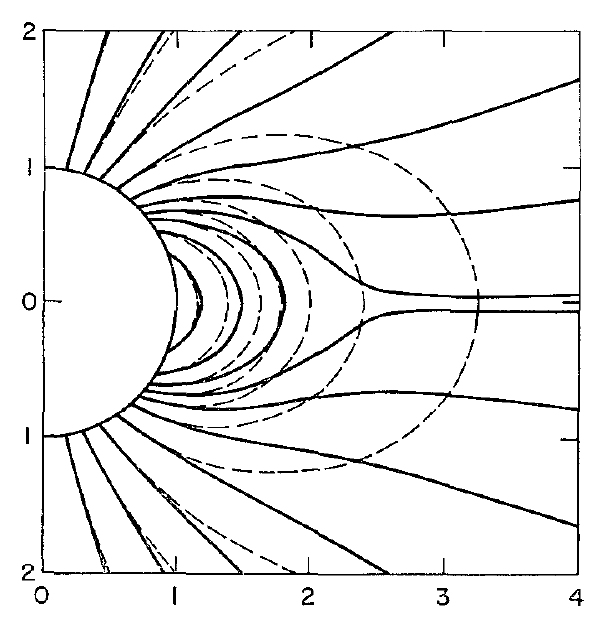}
\caption{A comparison of the magnetic field (solid curves) with a potential field (dashed curves) having the same normal component at the reference level. The field lines for the two configurations are chosen so as to be coincident at the surface. \citep{Pneuman:1971a}.
\label{fig:pneuman}
}
\end{figure}

Early numerical models prescribed volumetric coronal heating in ways that strived to accounted for the effects thermal conduction and radiative losses,  as well as satisfying known constraints of coronal heating. However, these works found that the observed fast solar wind (speed 700-800 km/s) cannot  be produced by thermal pressure without temperatures greatly exceeding  coronal values, particularly in coronal holes where the fast wind originates.  

Using simplified conservation laws, \cite{Pneuman:1966a, Pneuman:1968a, Pneuman:1969a} obtained helmet-like closed field line regions buffeted by converging open field lines (\figurename~\ref{fig:pneuman_helmet}). Later, \cite{Pneuman:1971a} numerically solved the conservation equations for an axially symmetric steadily expanding corona with an embedded magnetic dipole, assuming North-South symmetry. In this case the solution is not only axially symmetric, but it is also symmetric with respect to the equatorial plane. 

The \cite{Pneuman:1971a} solution shows several interesting features. An important aspect of the solution is the development of a neutral point at the top of the helmet where the magnetic field vanishes. A current sheet forms between the regions of opposite magnetic polarity above the neutral point. The outflowing plasma reaches the \alf velocity just above the neutral point (at the top of the helmet), so the top of the helmet is, in effect, the transition from sub-Alfv{\'e}nic to super-Alfv{\'e}nic flow.

\figurename~\ref{fig:pneuman} shows a comparison of the field lines of the numerical solution and a potential field model with the same normal component at the reference level \citep{Pneuman:1971a}. To make the comparison meaningful, the field lines for the two cases are chosen so as to be coincident at the surface. As expected, the field is everywhere stretched outward by the gas with this distention becoming large near the neutral point. In the closed region well below the cusp the difference between the two configurations is small, mainly because the pressure at the reference level is taken to be independent of latitude. The outward distention of the field in this region, as a result, is produced solely by currents generated by the expansion along open field lines. For the general case of variable surface pressure, however, a pressure and gravitational force balance cannot be satisfied normal to the field and significant ${\bf j}\times{\bf B}$ forces are expected to maintain the equilibrium.

In spite of its obvious limitations, the \cite{Pneuman:1971a} simulation established the usefulness of MHD simulations of the solar corona. It was the first successful attempt to apply the conservation laws of magnetohydrodynamics to explain large-scale features of the solar corona.

\section{Steady-state models of the solar wind} 
\label{sec:steadywind}

We adopt the ``ambient-transient'' paradigm to modeling the dynamic and highly structured solar wind. By ``ambient'' we mean a quiet-sun-driven full 3-D structure for the interplanetary magnetic field and a 3-D distribution of the solar wind parameters. They are both close to being steady-state in the frame of reference co-rotating with the Sun, except for the highly intermittent solar wind region.  This ambient solution determines the bimodal structure of the solar wind. It affects magnetic connectivity between the active regions at the Sun and  the corresponding regions in the heliosphere. In turn such connectivity affects energetic particle acceleration and transport.  

As we discussed in Section~\ref{sec:first_corona}, the first generation of magnetohydrodynamic models of the interplanetary medium were developed in the second half of the 1970s and were used for about two decades \citep{Steinolfson:1975a, Steinolfson:1978a, Pizzo:1978a, Pizzo:1980a, Pizzo:1982a, Steinolfson:1982a, Steinolfson:1988a, Steinolfson:1988b, Pizzo:1989a, Steinolfson:1990a, Pizzo:1991a, Steinolfson:1992a, Pizzo:1993a, Pizzo:1994a, Pizzo:1994c, Pizzo:1994b, Steinolfson:1994c}. These models were designed to describe only large-scale bulk-average features of the plasma observed through the solar cycle. At solar minimum, these coronal structures are the following: 
\begin{enumerate}
\item
open magnetic field lines forming coronal holes;  
\item
closed magnetic field lines forming a streamer belt at low latitudes; 
\item
the bimodal nature of the solar wind is reproduced with fast wind  originating from coronal holes over the poles and slow wind at low latitudes.  
\end{enumerate}
A thin current sheet forms at the tip of the streamer belt and separates opposite directed magnetic flux originating from the two poles.  At solar minimum, the fast wind lies at  30 degrees heliographic latitude and has an average velocity of  750 km s$^{-1}$ at distances greater than 15 solar radii, at which distance the wind has attained the majority of its terminal velocity.  The slow wind, by contrast, is confined close to the global heliospheric current sheet, which lies near the equator at solar minimum.  This component of the wind is highly variable, with speeds that lie between 300 and 450 km s$^{-1}$.  The slow solar wind has been suggested by \cite{Wang:1990a} to originate from highly expanding plasma traveling down magnetic flux tubes that originate near coronal hole boundaries.  It has also been suggested that opening of closed flux tubes by interchange reconnection with open flux may release plasma to form the slow solar wind \citep{Fisk:1998a, Lionello:2005a, Rappazzo:2012a}. More recent theories have related the slow solar wind to complex magnetic topology flux tubes near the heliospheric current sheet, which are characterized by the squashing factor \citep{Titov:2012a, Antiochos:2012a}. At solar maximum, the current sheet is highly inclined with smaller coronal holes forming at all latitudes, while the fast wind is largely absent.  



\subsection{2D and quasi-3D models}
\label{sec:ideal}

\cite{Steinolfson:1975a} numerically solved the MHD equations for a spherically symmetric (2D) solar corona, neglecting the polar components of velocity and magnetic field. They solved for $\rho$, $v_r$, $v_\phi$, $p$, $B_r$, and $B_\phi$ as a function of time and radial distance. The model was applied to a forward-reverse shock pair propagating in this simplified solar wind and the solution was compared with the results of similarity theory. In the end \cite{Steinolfson:1975a} concluded that MHD effects might be important in the dynamical behavior of the solar wind near Earth orbit.

\begin{figure}[htb]
\centering
\includegraphics[width=0.6\textwidth]{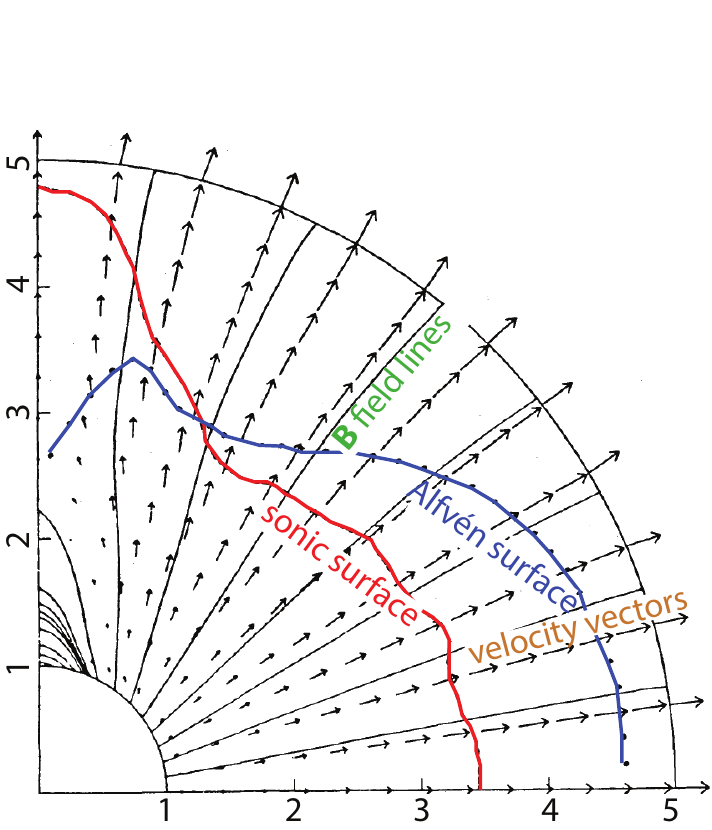}
\caption{Steady-state coronal structure for a plasma $\beta$ value of 0.5. Solid lines represent magnetic field lines, while arrows show plasma flow velocity vectors. The horizontal axis is the solar equator \citep{Steinolfson:1982a}.
\label{fig:steinolfson1}
}
\end{figure}

In a follow-up paper, \cite{Steinolfson:1978a} considered a different situation, when the flow and magnetic field are in the meridional plane. They solved time evolution equations for $\rho$, $v_r$, $v_\theta$, $p$, $B_r$, and $B_\theta$ as a function of polar angle and radial distance. They considered solar transients in two magnetic configurations: with the magnetic field radial (open), and with the magnetic field parallel to the solar surface (closed). The solar event is simulated by a pressure pulse at the base of an initially hydrostatic atmosphere. The pressure pulse ejects material into the low corona and produces a disturbance that propagates radially and laterally through the corona. The disturbance is preceded by waves (which may strengthen into shocks) that travel in the meridional plane with the shape of an expanding loop. The portion of the disturbance between the ejected material and the preceding waves consists entirely of coronal material whose properties have been altered by the waves. This simulation, in spite of its many limitations, was the first successful attempt to simulate coronal transients.

A few years later, \cite{Steinolfson:1982a} revisited the steady global solar corona that was investigated a decade earlier by \cite{Pneuman:1971a}. This more sophisticated simulation allowed for non-constant temperature and did not require specific conditions to be met at the cusp \citep{Pneuman:1971a}. They applied their previous model \citep{Steinolfson:1978a} to study the axially symmetric global corona. Their solution essentially confirmed the configuration obtained by \cite{Pneuman:1971a} with some notable differences. Since the coronal plasma is subsonic and sub-Alfv{\'e}nic near the Sun the flow is considerably more complex than predicted by \cite{Pneuman:1971a}. The main feature, however, is essentially the same: the coronal outflow results in closed field lines near the equator and open field lines (coronal holes) at high latitudes (see \figurename~\ref{fig:steinolfson1}).

In parallel with Steinolfson's efforts, Pizzo took a very different approach to simulating the interplanetary medium \citep{Pizzo:1978a, Pizzo:1980a, Pizzo:1982a, Pizzo:1989a, Pizzo:1991a, Pizzo:1993a, Pizzo:1994a, Pizzo:1994b, Pizzo:1994c}. As a first step, \cite{Pizzo:1978a} developed a 3D hydrodynamic model of steady corotating streams in the solar wind, assuming a supersonic, inviscid and polytropic flow beyond approximately $35\,R_s$. This approach takes advantage of the fact that in the inertial frame the temporal and azimuthal gradients are related by $\partial_t = - \Omega_\sSun \partial_\phi$, where $\phi$ is the azimuth angle and $\Omega_\sSun$ is the equatorial angular velocity of the Sun.

The steady-state conservation equations were solved using an explicit Eulerian approach on a rectangular ($\theta,\phi$) grid covering the $45^\circ\le\theta\le135^\circ$ and $-60^\circ\le\phi\le60^\circ$ spherical surface. Periodic boundary conditions were imposed at the azimuthal edges of the mesh, while the latitudinal boundaries are free surfaces, with the meridional derivatives approximated by one-sided differences. The density, velocity vector, and scalar pressure were defined at the inner boundary ($35\,R_\sSun$). Knowing the solution at a heliocentric distance $r$, the solution was advanced to $r+\Delta r$ using the conservation equations. Using a value of $\Delta r=30$km, \cite{Pizzo:1978a} obtained a steadily corotating stream structure between $35\,R_\sSun$ and 1AU.

Specifically, \cite{Pizzo:1978a, Pizzo:1980a} solved the governing equations that describe the dynamical evolution
of 3-D corotating solar wind structures. The model is limited to those structures that are steady or nearly steady in the frame rotating with the Sun and utilizes the single-fluid, polytropic, nonlinear, 3D hydrodynamic equations to
approximate the dynamics that occur in interplanetary space, where the flow is supersonic and the
governing equations are hyperbolic. In the inertial frame, the equations are the following  \citep{Pizzo:1978a, Pizzo:1980a}:
\begin{equation}
  - \Omega_\sSun \frac{\partial \rho}{\partial \phi} + {\bf\nabla}\cdot \left(\rho {\bf u}\right) = 0
\label{eq:pizzo1}
\end{equation}
\begin{equation}
  - \Omega_\sSun \left({\bf e}_r \frac{\partial u_r}{\partial \phi}
  + {\bf e}_\theta \frac{\partial u_\theta}{\partial \phi}
  +{\bf e}_\phi \frac{\partial u_\phi}{\partial \phi}\right)
   + \left({\bf u}\cdot{\bf\nabla}\right) {\bf u} = 
   - \frac{1}{\rho} {\bf\nabla} p - \frac{G M_\sSun}{r^2} {\bf e}_r
\label{eq:pizzo2}
\end{equation}
\begin{equation}
  \left(- \Omega_\sSun \frac{\partial}{\partial \phi} + {\bf u}\cdot{\bf\nabla} \right) \left(\frac{p}{\rho^\gamma}\right)= 0 \,,
\label{eq:pizzo3}
\end{equation}
where $\rho$ is the mass density, ${\bf u}$ is the center of mass velocity, $p$ is the total isotropic (scalar) gas pressure, $G$ is the gravitational constant, $M_\sSun$ is the solar mass, $\gamma$ is the polytropic index and $\Omega_\sSun$ is the equatorial angular rotation rate of the Sun. The independent variables are the spherical polar coordinates $(r,\theta,\phi$). Conduction, wave dissipation, differential rotation, the magnetic field, and shock heating are all neglected. Equations~(\ref{eq:pizzo1}--\ref{eq:pizzo3}) can be rearranged to obtain a set of differential equations describing the radial evolution of the primitive variables and solved by marching in radial distance from the inner boundary outward. This method yields a 3D solution that is steady-state in the corotating frame and describes the stream structure in the interplanetary medium.

\begin{figure}[htbp]
\centering
\includegraphics[width=0.6\textwidth]{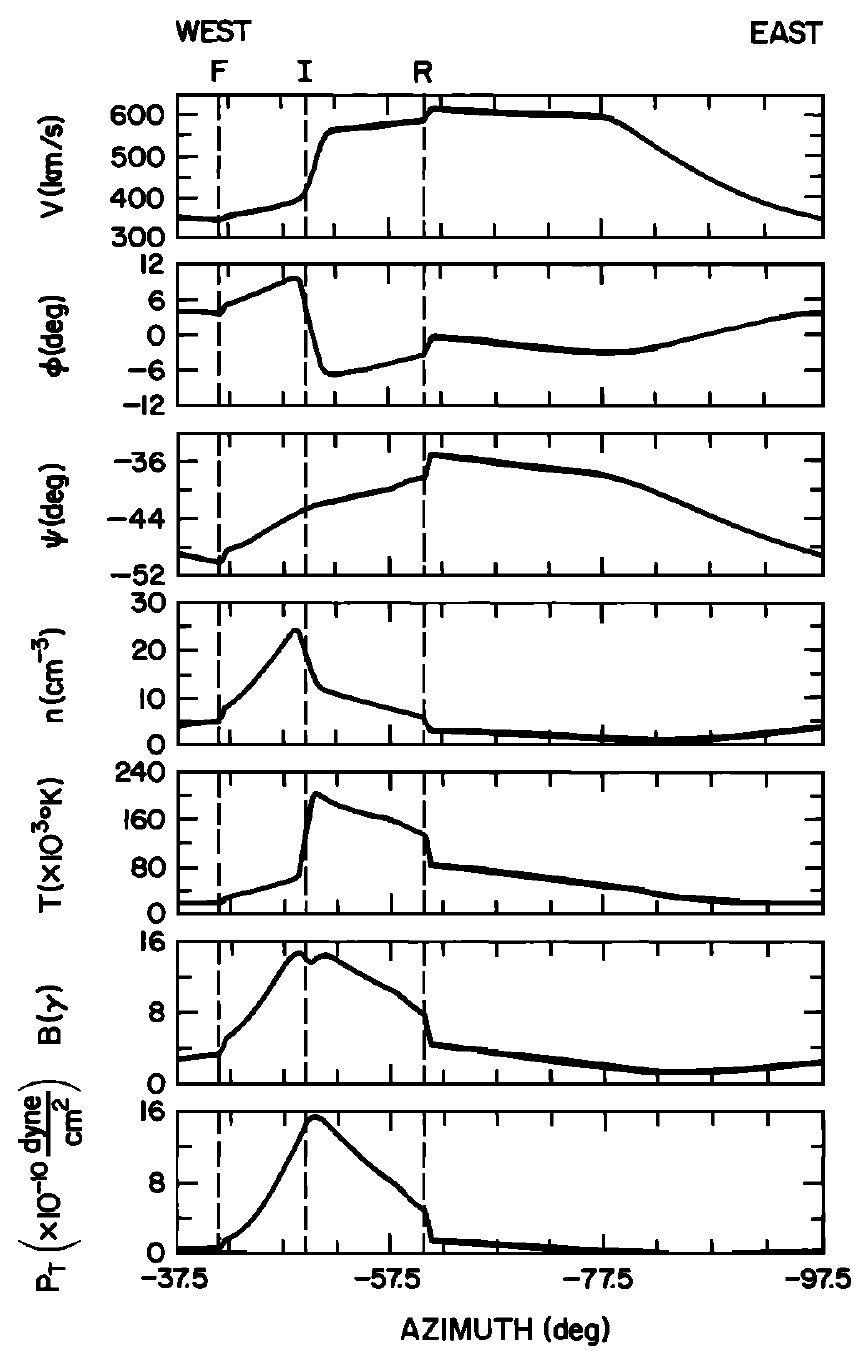}
\caption{Equatorial solution for an initially steep-sided circular stream at I AU. F and R mark the forward and reverse wave fronts, which demarcate the dynamic compression ridge. The two shocks propagate in opposite directions, but both are convected outward in the bulk flow. The interface, I, is a shear layer separating fast and slow flow
regimes that initially had very different fluid properties \citep{Pizzo:1982a}.}
\label{fig:pizzo1}
\end{figure}

In a subsequent paper, \cite{Pizzo:1982a} extended his marching method to include the interplanetary magnetic field. He considered the implication of a high-speed stream emanating from the Sun and expanding to 1~AU. The central portion of the stream is a circular plateau some 30$^\circ$ in diameter where the radial velocity is a uniform 600 km/s. The speed falls off smoothly in all directions, bottoming out at the background speed of 300 km/s at a distance of 7.5$^\circ$ from the periphery of the plateau. The results at 1~AU are shown in \figurename~\ref{fig:pizzo1}. The panels from top to bottom show the bulk speed, the flow angle in the inertial frame, the flow angle in the rotating frame, the number density, the single-fluid temperature, the magnetic field intensity and the total pressure as a function of azimuth. At the front of the stream (west, or left), forward (F) and reverse (R) waves have formed about an interface (I). The two waves, which propagate in opposite directions from the interface, originate in the general compression at the stream front near the Sun.

The \cite{Pizzo:1982a} model was later applied to qualitatively explain Ulysses observations at larger helio-latitudes \citep{Pizzo:1994a}. Ulysses discovered that the forward-reverse shock pair structure commonly bordering corotating interaction regions beyond 1~AU  near the ecliptic plane undergoes a profound change near the maximum heliographic latitude of the heliospheric current sheet.  At this latitude the forward shock is observed to weaken abruptly, appearing as a  broad forward wave, while the  reverse shock weakens much more slowly \citep{Gosling:1993b}. These observations were well reproduced by the \cite{Pizzo:1982a} model \citep{Pizzo:1994a}.  

\subsection{Connecting the corona and the heliosphere}
\label{sec:WSA}

Probably the most challenging region to model is the transition from the cold ($\sim5000$ K) photosphere to the million K solar corona. This narrow, but critical region has very complex physics and challenging numerics: the plasma temperature increases by a factor of several hundred over $\sim500$ km ($<10^{-3}$$R_\sSun$). The first attempts to include this physics in 3D simulations are just beginning \citep[see Section \ref{sec:igor2} in this paper and][]{Sokolov:2016a}.

In the early 1990s it was recognized that the solar wind speed at 1 AU negatively correlates with a magnetic flux tube expansion factor near the Sun \citep{Wang:1990a, Wang:1992a, Wang:1995b}. This expansion factor describes the ratio of a given flux tube's cross sectional area at some heliocentric distance and the cross sectional area of the same flux tube at the solar surface. \cite{Wang:1990a, Wang:1992a, Wang:1995b} found that the solar wind speed at a heliocentric distance can be expressed as 
\begin{equation}
  u(\mathbf{r}) = u_{min} + \frac{u_{max}-u_{min}}{f_{exp}(\mathbf{r})^\alpha} \,,
\label{eq:ws1}
\end{equation}
where $u_{min}$ and $u_{max}$ represent the slow and fast solar wind speeds, $f_{exp}$ is the expansion factor at heliocentric location $\mathbf{r}$, while $\alpha$ is an empirical factor (near unity). Later \cite{Arge:2000a} and \cite{Arge:2004a} generalized the \cite{Wang:1990a, Wang:1992a, Wang:1995b} formula and introduced several additional parameters. A comprehensive description of the Wang-Sheeley-Arge (WSA) model and its parameter values can be found in \cite{Riley:2015a}.

The WSA formula provides an efficient and simple way to circumvent the complex physics of the low corona and transition region. It is used by a number of heliosphere models to provide inner boundary conditions beyond the \alf surface (where the solar wind speed exceeds the local \alf speed). These models typically place their inner boundaries in the $20$ to $30$ $R_\sSun$ range \citep[\eg][]{Wold:2018a}.

\subsection{3D MHD heliosphere models}
\label{sec:3D-MHD}

Pizzo's early work was followed with the development of the ENLIL heliospheric model by \cite{Odstrcil:1999a}, who applied it to study several phenomena \citep{Odstrcil:1996a, Odstrcil:1997a} including coronal mass ejection (CME) propagation \citep{Odstrcil:1999a, Odstrcil:2004_chip_a, Odstrcil:2005_chip_a, Siscoe:2008_chip_a}. ENLIL has been adapted to accept inner boundary solar wind conditions from a variety of sources, including the WSA model \citep{Wang:1990a, Arge:2004a} and coronal MHD models \citep{Odstrcil:2002a, Odstrcil:2004a}.  In the case of \cite{Hayashi:2012a}, the inner boundary conditions (outside the critical point) are derived from interplanetary scintillation (IPS) observations \citep{Jackson:1998_chip_a}. More recently, other groups also developed 3D inner heliosphere models with super-Alfv{\'e}nic inner boundary conditions using the WSA approach. These include the LFM-Helio model \citep{Merkin:2011a, Merkin:2016b}, the SUSANOO-SW code \citep{Shiota:2014a, Shiota:2016_chip_a}, the MS-FLUKSS suite \citep{Kim:2016a} and EUHFORIA developed by the Leuwen group \citep{Pomoell:2018a}.

\begin{figure}[htb]
\centering
\includegraphics[width=0.7\textwidth]{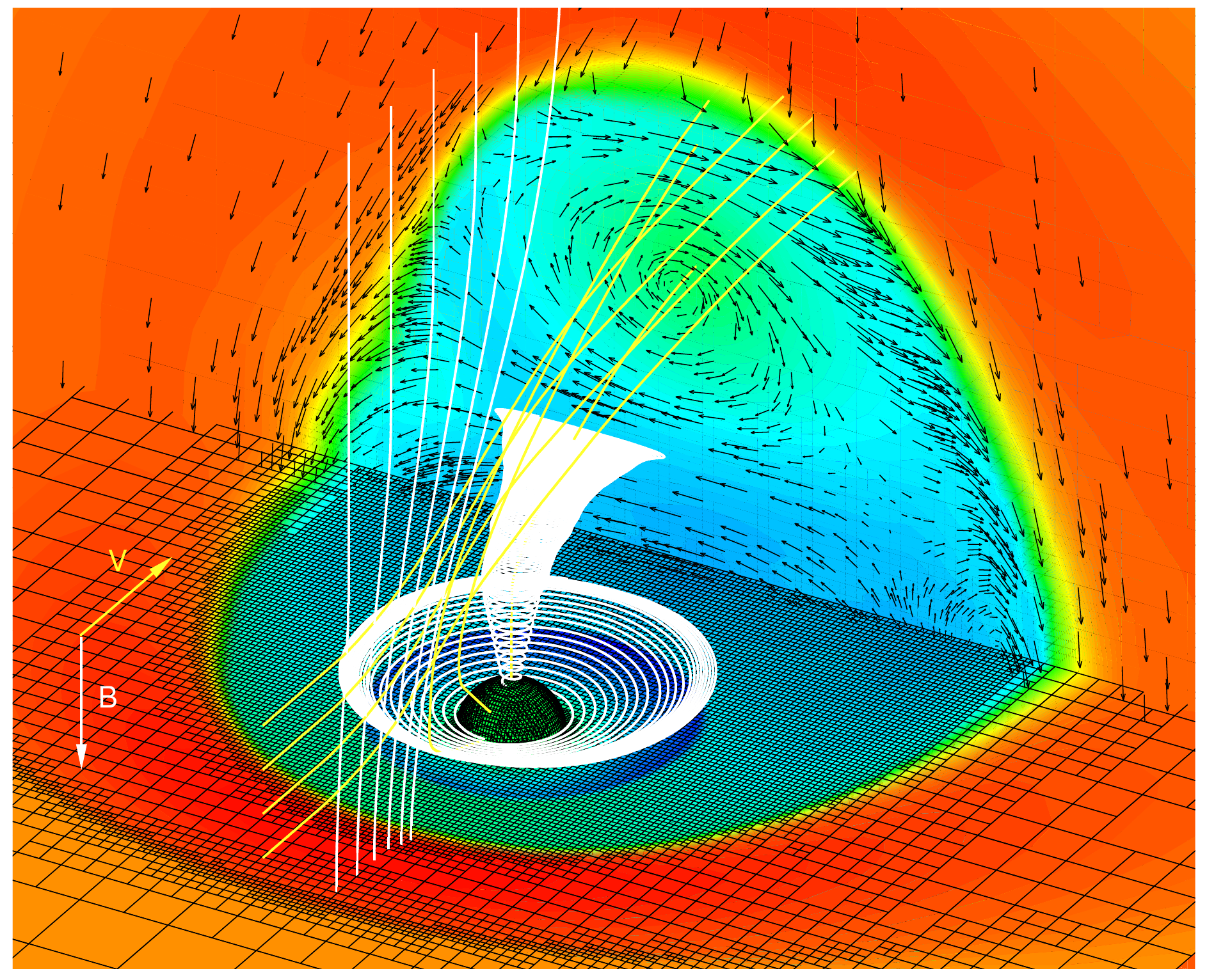}
\caption{Three-dimensional view of the global heliosphere. The color code shows the log of plasma density. Yellow lines are the plasma velocity streamlines and the white lines follow the magnetic field lines. Black arrows indicate the direction of the magnetic field along a cross-tail cut placed 225 AU downstream from the Sun. \citep[from][]{Linde:1998a}}
\label{fig:linde}
\end{figure}

WSA-like empirical inner boundary conditions were also used in the first generation of outer heliosphere models describing the interaction between the solar wind and the interstellar medium. \cite{Linde:1998a} published the first 3D MHD model describing the interaction of the magnetized solar wind with the magnetized interstellar medium. This simulation also took into account the presence of the neutral component of the interstellar medium and the resulting mass loading process inside the heliosphere. \figurename~\ref{fig:linde} presents a 3D view of the global heliosphere. 

Following the early work of \cite{Linde:1998a}, several groups developed sophisticated 3D models of the outer heliosphere and its interaction with the magnetized interstellar medium. While details of these models go beyond the scope of this review we note the progress made by \cite{Opher:2003a, Opher:2006a, Opher:2007a, Opher:2016a} and the Huntsville group \citep{Pogorelov:2013a, Pogorelov:2015a, Zirnstein:2017a, Pogorelov:2017a}.


\subsection{3D MHD coronal models}
\label{sec:3D-corona}

The next level of sophistication of MHD coronal models came with reduced \citep{Usmanov:1993a, Mikic:1999a} and variable adiabatic index models \cite[\eg][]{Wu:1999a, Roussev:2003a, Cohen:2007a}.  These models approximate coronal heating by greatly lowering the adiabatic index from  ${5}/{3}$ to typically 1.05 so that the plasma remains nearly isothermal  as it expands and maintains the pressure necessary to drive a fast wind.  Such models have been successful in largely reproducing the observed solar wind speed at 1~AU over an entire solar cycle.

A number of coronal heating models have been published over the years using two fundamentally different approaches: (i) use a general heating function with an \textit{ad-hoc} heating rate that is chosen to fit observations; or (ii) include a semi-empirical coronal heating function that is based on the physics of \alf waves. Examples for the first approach include papers by \cite{Groth:1999b, Groth:2000b, Lionello:2001a, Lionello:2009a, Riley:2006a, Feng:2007a, Nakamizo:2009a, Feng:2010a, Titov:2008a}, and \cite{Downs:2010a}.  An important limitation of this approach is that models utilizing an \textit{ad-hoc} approach depend on some free parameters that need to be determined for various solar conditions. While the \textit{ad-hoc} heating function approach is well-suited for typical conditions, it can't properly account for unusual conditions, such as those that can occur during extreme solar events.

A major benefit of the WSA model is its simplicity and therefore it can be seamlessly integrated into global-scale space weather simulations \cite[\eg][]{Odstrcil:2003a, Cohen:2007a}. In the \cite{Cohen:2007a} study the WSA formulae were used as the boundary condition for a large-scale 3-D MHD simulation with varied polytropic gas index distribution \citep[see][]{Roussev:2003a}. These models can successfully reproduce observed solar wind parameters at 1~AU. 

\begin{figure}[htb]
\centering
\includegraphics[width=0.8\textwidth]{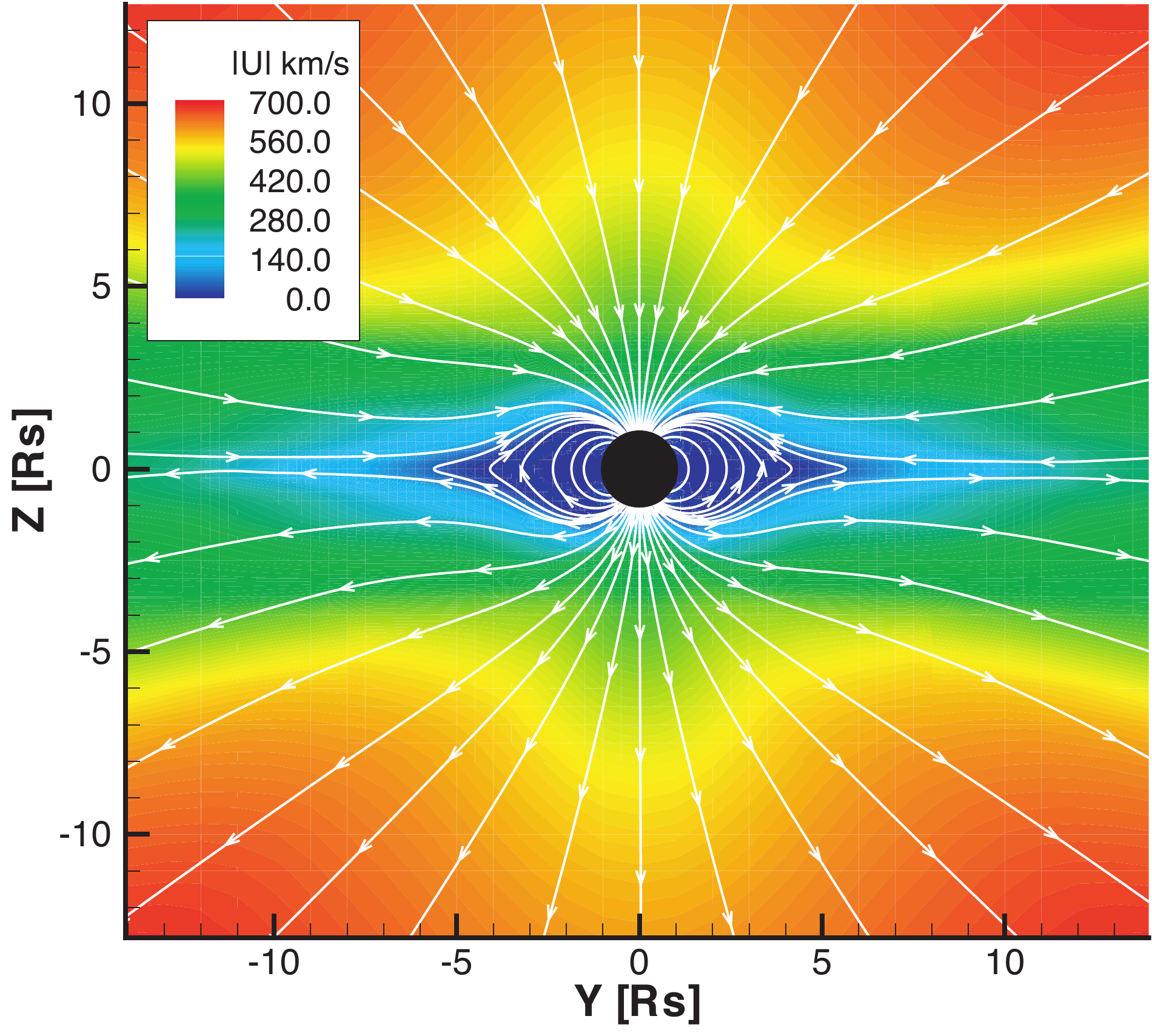}
\caption{Magnetic structure and velocity for the steady-state solar
wind solution by \cite{Manchester:2004a}.  Solid white lines are magnetic ``streamlines'' drawn
in the $y-z$ plane superimposed upon a false color image of
the velocity magnitude. Note the bimodal nature of the solar wind.}
\label{fig:wind_example}
\end{figure}

An example of a solar model driven by empirical heating is shown in Fig.~\ref{fig:wind_example}, which depicts solar minimum conditions on a meridional plane close to the Sun.  This result from \cite{Manchester:2004a}
 is based on a model by \cite{Groth:2000b}.  Here the color image indicates the velocity magnitude, $|{\bf u}|$, of the plasma while the magnetic field is represented by solid white lines.  Inspection reveals a bimodal outflow pattern with slow wind leaving the Sun below 400 km/s near the equator and high-speed wind above 700 km/s found above 30$^\circ$ latitude.  In this model, we find that the source of the slow solar wind is plasma originating from the coronal hole boundaries that over-expands and fails to accelerate to high speed as it fills the volume of space radially above the streamer belt.  This model is consistent with the empirical model of the solar wind proposed by \cite{Wang:1994a} that explains solar wind speeds as being inversely related to the expansion of contained magnetic flux tubes.  The magnetic field remains closed at low latitude close to the Sun, forming a streamer belt.  At high latitude, the magnetic field is carried out with the solar wind to achieve an open configuration.  Closer to the equator, closed loops are drawn out and, at a distance $(r > 3 R_\sSun)$, collapse into a field reversal layer.  The resulting field configuration has a neutral line and a current sheet originating at the tip of the streamer belt.

Several validation and comparison studies have been published using this approach \citep{Owens:2008a, Vasquez:2008a, MacNeice:2009a, Norquist:2010a, Gressl:2014a, Jian:2015a, Reiss:2016a}. However, these models do not capture the physics of \alf wave turbulence or even neglect it altogether. Even though some of these models were designed to account for the \alf wave physics \cite[\eg][]{Cohen:2007a}, they do not capture many aspects of the interaction of the turbulence with the background plasma flow, which include both energy and momentum transfer from the turbulence to the solar wind plasma. Because Alfv{\'e}nic turbulence effects are likely to be of great significance in the near-Sun domain, these simpler models should be used with caution for simulating the solar atmosphere.

\subsection{Thermodynamic corona models}
\label{sec:thermo}

While the polytropic solar corona models were quite successful in describing the quiet low latitude corona they had major problems in accounting for the two-state (slow and fast) solar wind and large dynamical processes (such as CMEs) that can interrupt the quasi-steady state (in the corotating frame) situation. In particular, the shock thermodynamics got seriously distorted by the reduced adiabatic index and various spurious phenomena (like runaway plasma temperatures) appeared in the solutions.

\begin{figure}[htb]
\centering
\includegraphics[width=0.7\textwidth]{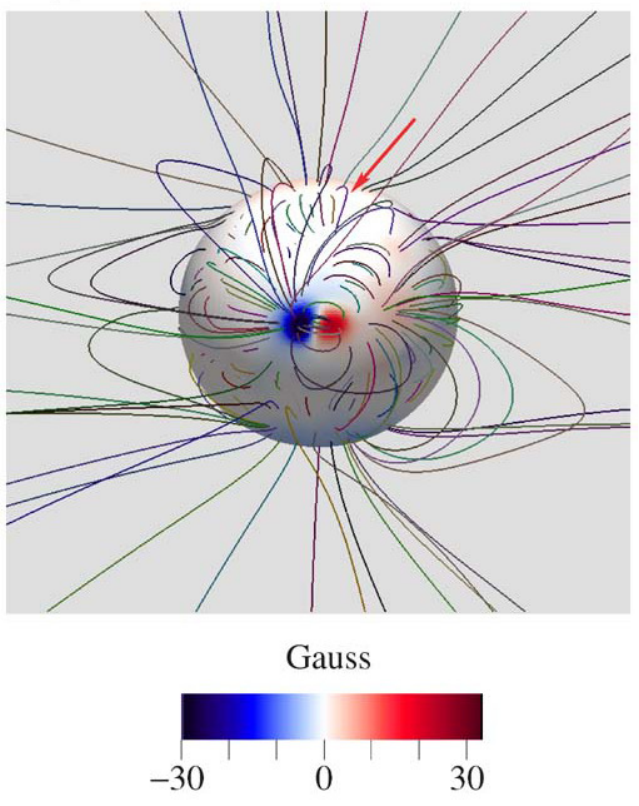}
\caption{Magnetic field model for the Sun around the time of Whole Sun Month (Carrington Rotation 1913, 1996 August 22 to September 18). The simulation was carried out with the thermodynamic model described in Section~\ref{sec:thermo}. The magnetic flux distribution was projected on the solar surface with selected magnetic field lines from the MHD solution. \cite[from][]{Lionello:2009a}}
\label{fig:lionello}
\end{figure}

In order to overcome this problem the  full energy equation -- with all accompanying computational challenges -- needs to be considered. \cite{Mikic:1999a} proposed that the energy equation be solved with a realistic adiabatic index ($\gamma=5/3$) and an empirical heating function be introduced to account for the effects of heat conduction, radiative cooling, and various heating processes. In particular, \cite{Mikic:1999a} introduced the following form of the heating function:
\begin{equation}
  S = -{\bf\nabla}\cdot{\bf q} - n_e n_p Q(T) + H_{ch} + H_d + D \,,
\label{eq:cor-heat1}
\end{equation}
where $H_{ch}$ is the coronal heating source, $D$ is the \alf wave dissipation term, $H_d =\eta J^2 + \nu {\bf\nabla} {\bf v}\colon {\bf\nabla} {\bf v}$ represents heating due to viscous and resistive dissipation, and $Q(T)$ is the radiative loss function. In the collisional regime (below $\sim10R_\sSun$), the heat flux is ${\bf q}= {\bf b}({\bf b}\cdot{\bf\nabla})T$, where ${\bf b}$ is the unit vector along the magnetic field vector, and $\kappa_{\parallel} = 9\times10^7\,T^{5/2}$ is the Spitzer value of the parallel thermal conductivity. The polytropic index $\gamma$ is $5/3$. In the collisionless regime (beyond $\sim10R_\sSun$), the heat flux is modeled by ${\bf q} = \alpha n_e k T {\bf v}$, where $\alpha$ is an empirical parameter. The coronal heating source is a parameterized function given in the form of
\begin{equation}
  H_{ch} (r,\theta) = H_{0} (\theta) \exp\left[-\frac{r-R_\sSun}{\lambda(\theta)}\right] \,,
\label{eq:cor-heat2}
\end{equation}
where the empirical functions $H_0(\theta)$ and $\lambda(\theta)$ express the latitudinal variation of the volumetric heating and scale length, respectively.

While this ``thermodynamic'' approach sidesteps the underlying physics of coronal heating and solar wind acceleration it provides an adequate mathematical framework to describe the coronal processes in a way that is consistent with solar dynamical processes.

The thermodynamic coronal model has been successfully applied to simulate the solar corona for the first whole Sun month \citep{Lionello:2009a}. The global magnetic configuration obtained with the model is shown in \figurename~\ref{fig:lionello}.

\subsection{Model inputs}
\label{sec:input}

\begin{figure}[htb]
\centering
\includegraphics[width=0.9\textwidth]{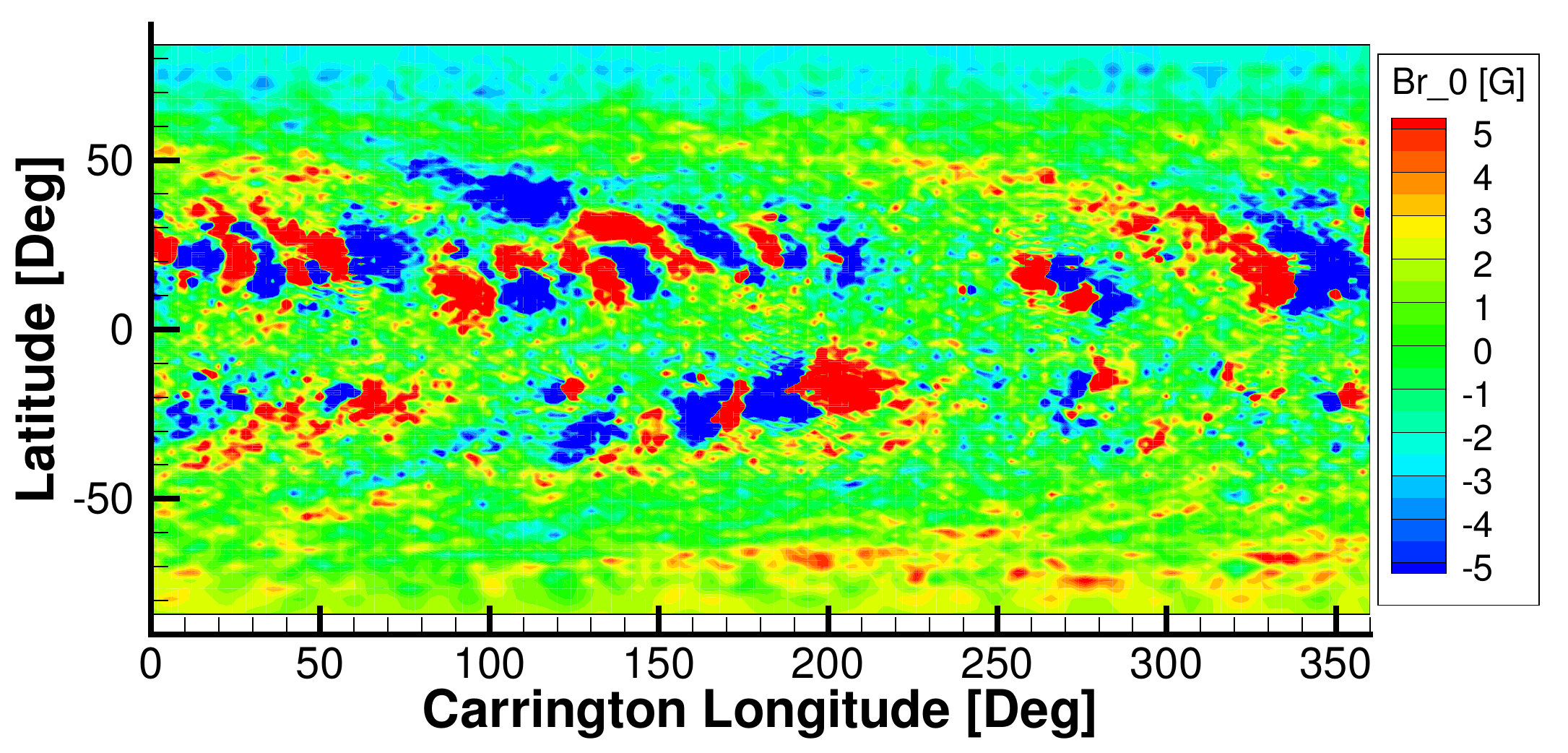}
\caption{Carrington map of the radial magnetic field component at $1\;R_\sSun$. This map is based on a synoptic magnetogram of Carrington rotation 2109 (2011 April 12 to May 9) from GONG and processed to a PFSS solution using spherical harmonics. For the purpose of showing both the coronal holes and active regions, the magnetic field in this plot is saturated by $\pm5\;$G.}
\label{fig:gong}
\end{figure}

The magnetic field is an essential component of the corona. As a matter of fact, the surface magnetic field distribution is the primary direct quantitative observable for models, physical variables associated with other observations need to be inferred. Early models used 
simple dipolar magnetic fields to simulate solar minimum conditions 
\citep{Groth:2000a, Steinolfson:1982a}.  In this case, the dipole is chosen 
such that the maximum field strength at the poles approximately 10~gauss.
However, simulating realistic solar conditions requires the use of 
the observed line-of-sight global magnetic field. To provide these data, 
full disk magnetograms are taken as the Sun completes a full rotation, and 
then combined into a full-surface synoptic map. Early examples of these
maps include those provided by the Stanford Wilcox and Mount Wilson observatories as well as the Global Oscillation Network Group (GONG). These data sources are further augmented by offerings from 
the National Solar Observatory's Synoptic Long-term Investigation of the
Sun (SOLIS), and NASA's Solar Dynamics Observatory/Helioseismic and 
Magnetic Imager (HMI). Figure \figurename~\ref{fig:gong} provides an example input of a full-surface synoptic map based on GONG data. The use of synoptic maps in global MHD models 
was pioneered by \cite{Wu:1999a, Mikic:1999a, Linker:1999a}
and applied by many others \cite[\eg][]{Roussev:2003a, Hayashi:2005a,
Feng:2007a, Cohen:2007a, Nakamizo:2009a, vanderholst:2010a}.

While the line-of-sight photospheric magnetic field can be accurately 
measured and extrapolated to coronal heights with reasonable accuracy, 
the base densities and temperatures remain much more difficult to 
accertain, especially on the global scale required to specify boundary
conditions for numerical models.  In the case of \cite{vanderholst:2010a},
the base temperature and mass density were derived empirically from the
differential emission measure tomography (DEMT) technique by 
\cite{Vasquez:2008a}. While promising, this approach requires 
an involved, time-consuming calculation not suitable for operational
use. In recent years, a physics-based approach has been used to 
self-consistently calculate the thermodynamic properties at the base
of the corona by applying field-aligned electron heat conduction and
radiative processes (typically derived from Chianti \citep{Dere:1997a}). 
With this approach, it is possible to self-consistently reproduce the 
transition region. With that region, and the appropriate coronal density and 
temperatures, realistic heating functions can be applied
\citep{Lionello:2011a, Sokolov:2013a,vanderholst:2014a}.

\subsection{Model validation}
\label{sec:predict}

\begin{figure}[htb]
\centering
\includegraphics[width=0.9\textwidth]{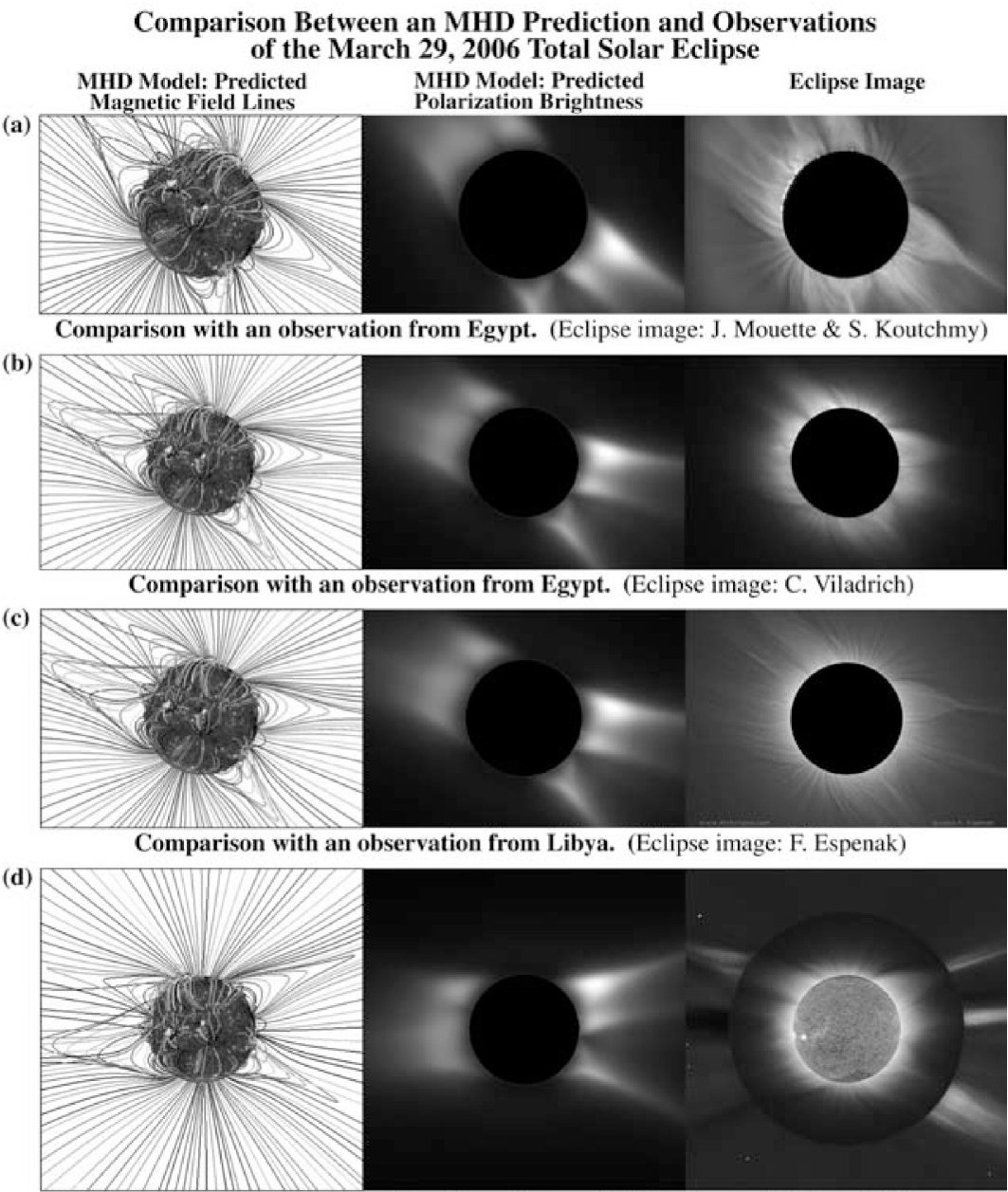}
\caption{Comparison between the MHD prediction (with magnetic field lines and polarization brightness shown in the first two columns) and eclipse observations (shown in the third column). \citep{Mikic:2007a}.}
\label{fig:mikic1}
\end{figure}

\begin{figure}[htb]
\centering
\includegraphics[width=\textwidth]{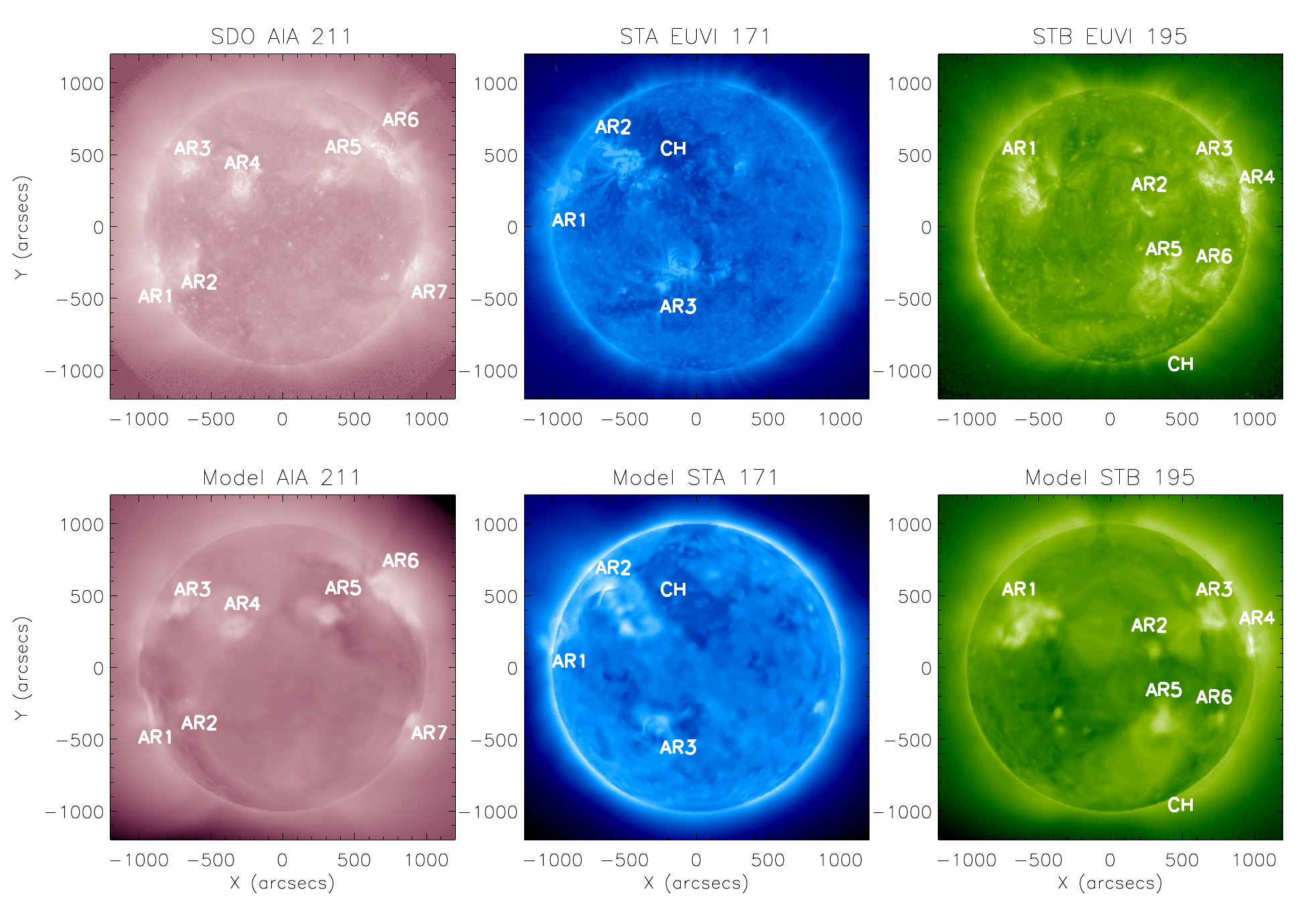}
\caption{Top panels (left to right): observational
images from SDO/AIA 211{\AA}, STEREO A/EUVI 171{\AA},
and STEREO B/EUVI   195{\AA}. The observation time is 2011 March
7 20:00 UT (CR2107). Bottom panels: synthesized EUV images of
the model. Active regions and coronal holes are marked in both the
observational and synthesized images, to demonstrate the reproducibility
of the observed morphological structures in our simulations. (From \cite{Jin:2017a})
\label{fig:EUV}}
\end{figure}

3D coronal models applying synoptic magnetograms, in conjunction 
with thermodynamic processes such as field-aligned heat conduction, are
capable of both predicting and interpreting the detailed magnetic and 
plasma structure of the corona.  Successful comparisons began with the predicted appearance of the 
corona in Thomson-scattered white light, compared to coronagraph and 
eclipse images \citep{Mikic:1999a}.  Figure \figurename~\ref{fig:mikic1} 
provides a very impressive prediction of white-light images of a solar 
eclipses \citep{Mikic:2007a}. Models capture the low-density coronal 
holes and high-density helmet streamers, including plasma sheets 
extending into the low corona.  Where models can reproduce coronal mass
density and electron temperature, they can predict thermal emission in
the extreme ultraviolet and X-ray spectrum, and line-of-sight integration
yields synthetic images that show reasonably good agreement with
observations, as seen in Figure \ref{fig:EUV}. 

When coronal models extend into the heliosphere, in particular beyond
1 AU, they offer predictions of the solar wind plasma parameters, including
charge state composition that can be compared directly to \textit{in-situ}
observations. Early examples include those by \cite{Wu:1999a}. Steady-state models can be validated by replicating solar wind observations for the synodic 
rotation period measured at the Earth, namely 27.27 days. Examples
are shown in \cite{Cohen:2008a, Feng:2012b, Meng:2015a, Toeroek:2018a}.


\subsection{Numerical mesh techniques}
\label{sec:numerics}


The computational domain for coronal simulations typically extends to
$r > 20 R_\sSun$. Flows will be superfast at the outer boundary
so that simple outflow boundary conditions will be well prescribed. 
Most simulations use spherical grids with fixed angular resolution, 
but highly stretched in the radial direction to provide high resolution
to the lower corona.  In the case of \cite{Groth:2000a} and \cite{Manchester:2004a}, 
an adaptive Cartesian grid was used with $4 \times 4 \times 4$ blocks, with
grid cells ranging in size from $1/32 R_\sSun$ to $2 R_\sSun$.  Grids 
are spatially positioned to highly resolve the corona as well as the 
heliospheric current sheet.  More recent models have used spheric adaptive
grids \citep[e.g.][]{Sokolov:2013a, vanderholst:2014a}, while others have 
employed a cube-sphere to maintain the advantage of nearly constant 
angular resolution while avoiding the singularity at the poles. 
For example, \cite{Feng:2007a, Feng:2010a, Feng:2012a} developed the 
Solar-InterPlanetary Conservation Element/Solution Element 
(SIP-CESE) MHD model that employs a six-component Yin-Yang grid 
system similar to a cubed sphere grid, with spherical shell-shaped 
domains extending from the low corona to 1~AU.

\section{Including \alf wave turbulence in  MHD Models} 
\label{sec:igor2}


\subsection{The role of \alf wave turbulence}
\label{sec:igor3}


The concept of \alf waves was introduced more than 70 years ago by Hannes \citet{Alfven:1942a}. The importance of the role they play in the heliosphere was not immediately recognized due to the lack of relevant observations. Results from Mariner~2 allowed an observational study of a wave-related phenomena in solar wind \citep{Coleman:1966a,Coleman:1967a}. This pioneering study culminated in a paper by \citet{Coleman:1968a}, a work that concluded that \alf wave turbulence has the potential to drive solar wind in a way that is consistent with observations at 1~AU. 

Interest in the role \alf waves play in the heating and acceleration of the solar wind goes back to the early years of \textit{in-situ} space exploration. Examples include papers by \cite{Belcher:1969a}, \cite{Belcher:1971a}, and by \cite{Alazraki:1971a}. A consistent and comprehensive theoretical description of \alf wave turbulence and its effect on the averaged plasma motion has been developed in a series of works, particularly by \citet{Dewar:1970a} and by \citet{Jacques:1977a, Jacques:1978a} (see also references therein). More recent efforts to simulate solar wind acceleration utilize the approach developed by \cite{Usmanov:2000a}. Currently, it is commonly accepted, that the gradient of the \alf wave pressure is the key driver for  solar wind acceleration, at least in fast flows. It is important to emphasize, that while incorporating the \alf wave-driven acceleration is usually accomplished by including the wave pressure gradient in the governing equations \citep{Jacques:1977a}, there is still no generally accepted approach to describe the coronal heating via the \alf wave turbulence cascade. 

\begin{figure}[htb]
\centering
\includegraphics[width=0.6\textwidth]{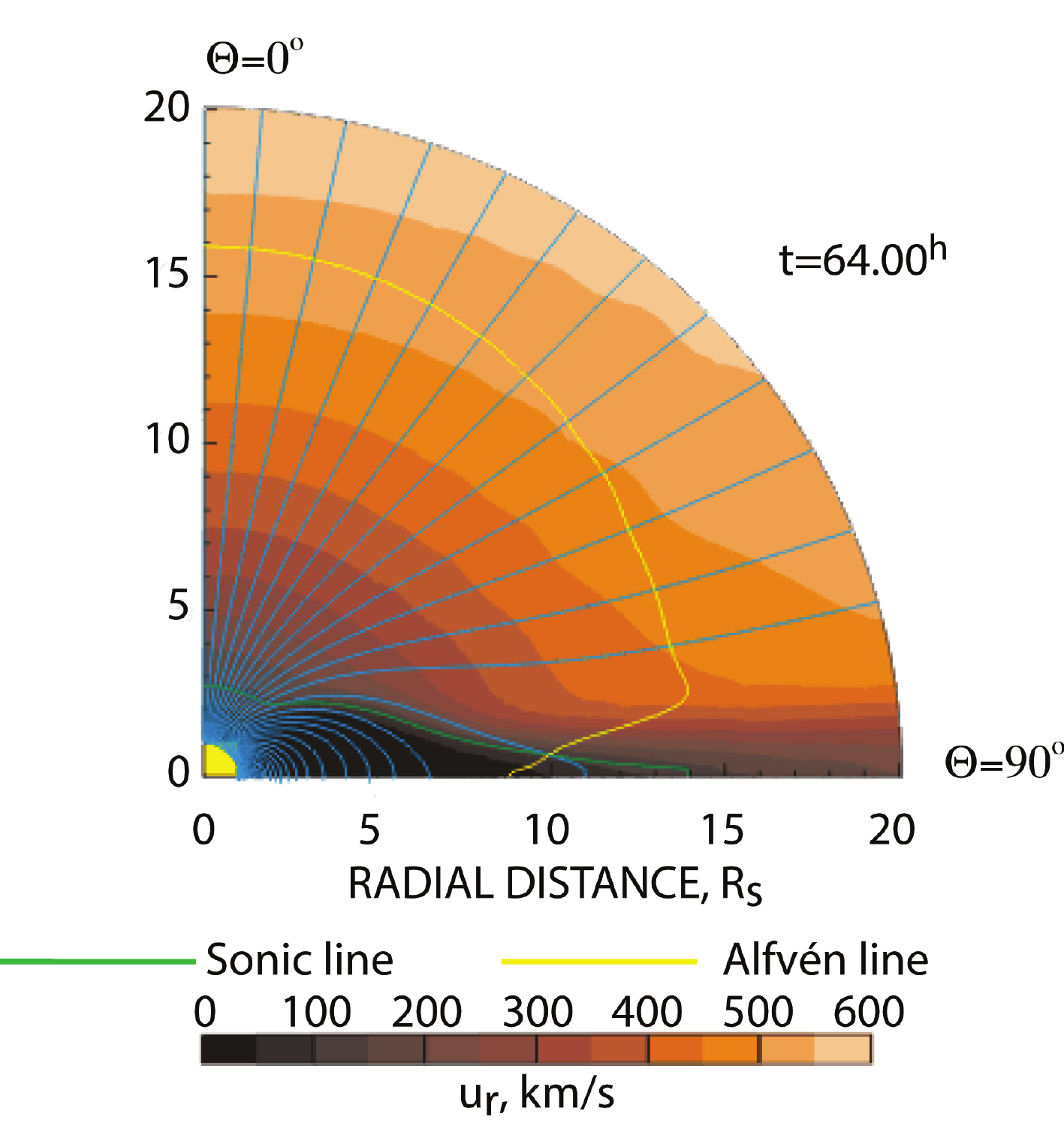}
\caption{Magnetic field configuration near the Sun superimposed on a map of radial flow velocities after 64 hours of relaxation. The \alf line (where the radial flow velocity is equal to the \alf velocity computed for the total magnetic field) is shown by the yellow line, and the sonic line (where the radial velocity is equal to the sound velocity) is shown by the green line. \citep{Usmanov:2000a}.
\label{fig:usmanov1}
}
\end{figure}

\figurename~\ref{fig:usmanov1} shows the results of the first axisymmetric (2-D) simulation of the solar corona and solar wind using self-consistent \alf turbulence \citep{Usmanov:2000a}. After 64 hours of relaxation time, a closed field region develops near the equator; the flow velocity is high in the
open polar field region but decreases toward the equator above the region of the closed magnetic field.

Damping of \alf wave turbulence as a source of coronal heating has also been extensively studied from the early days of \textit{in situ} solar wind observations \citep[\eg][]{Barnes:1966a,Barnes:1968a}.
Later, it was demonstrated that reflection from sharp pressure gradients in the solar wind \citep{Heinemann:1980a, Leroy:1980a} is a critical component of \alf wave turbulence damping  \citep{Matthaeus:1999a, Dmitruk:2002a, Verdini:2007a}.
For this reason, many numerical models explore the generation of reflected counter-propagating waves as the underlying cause of the turbulence energy cascade  \citep[\eg][]{Cranmer:2010a}, which transports the energy of turbulence from the large-scale motions across the \textit{inertial range} of the turbulence spatial scale to short-wavelength perturbations. The latter can be efficiently damped due to wave-particle interaction. In this way, the turbulence energy is converted to random (thermal) energy.

Many recent efforts aim to develop models that include \alf waves as a primary driving agent for both heating and  accelerating of the solar wind. Examples are  papers by \citet{Hu:2003a}, \citet{Suzuki:2005a}, \citet{Verdini:2010a}, \citet{Matsumoto:2012a}, and \citet{Lionello:2014a,Lionello:2014b}.

\subsection{Alfv{\'e}n wave turbulence driven solar atmosphere model}
\label{sec:igor4}

The \textit{ad hoc} elements can be eliminated from the solar corona model by assuming that the coronal plasma is heated by the dissipation of \alf wave turbulence \citep{Sokolov:2013a}. The dissipation itself is caused by the nonlinear interaction between oppositely propagating waves \cite[\eg][]{Hollweg:1986a}. 

Within coronal holes, there are no closed magnetic field lines, hence, there are no oppositely propagating waves.
Instead, a weak reflection of the outward propagating waves locally generates sunward propagating waves as quantified by \citet{vanderholst:2014a}. The small power in these locally generated (and almost immediately dissipated) inward propagating waves leads to a reduced turbulence dissipation rate in coronal holes, naturally resulting in the bimodal solar wind structure. Another consequence is that coronal holes look like cold black spots in the EUV and X-ray images,  while closed field regions are  hot and bright. Active regions, where the wave reflection is particularly strong, are the brightest in this model \citep[see][]{Sokolov:2013a, Oran:2013a, vanderholst:2014a}.

The continuity, induction, and momentum equations used in the model are the following:
\begin{equation}\label{eq:cont}
\frac{\partial\rho}{\partial t}+\nabla\cdot(\rho{\bf u})=0,
\end{equation}
\begin{equation}\label{eq:induction}
\frac{\partial{\bf B}}{\partial t}+\nabla\cdot\left({\bf u}{\bf B}-{\bf B}{\bf u}\right)=0,
\end{equation}
\begin{equation}\label{eq:momentum}
\frac{\partial(\rho{\bf u})}{\partial t}+\nabla\cdot\left(\rho{\bf u}{\bf u}-\frac{{\bf B}{\bf B}}{\mu_0}\right)+\nabla\left(p_i+p_e+\frac{B^2}{2\mu_0}+p_\sA\right)=-\frac{GM_\sSun \rho\,{\bf r}}{r^3} \,,
\end{equation}
where $\rho$ is the mass density, $\mathbf{u}$ is the bulk velocity ($u=|{\bf u}|$ is assumed to be the same for the ions and electrons), $\mathbf{B}$ is the magnetic field, $G$ is the gravitational constant, $M_\sSun$ is the solar mass, $\mathbf{r}$ is the position vector relative to the center of the Sun, $\mu_0$ is the magnetic permeability of vacuum. As has been shown by \citet{Jacques:1977a}, the \alf waves exert an \textit{isotropic} pressure ($p_\sA$ in the momentum equation). The relation between the wave pressure and wave energy density is $p_\sA=(w_++w_-)/2$.
 Here, $w_\pm$ are the energy densities for the turbulent waves propagating along the magnetic field vector ($w_+$) or in the opposite direction ($w_-$). The isotropic ion and electron pressures, $p_i$ and $p_e$, are governed by the appropriate energy equations:
\begin{eqnarray}\label{eq:energy}
\frac{\partial}{\partial
  t}\left(\frac{p_i}{\gamma-1}+\frac{\rho u^2}2+\frac{{\bf B}^2}{2\mu_0}\right)+\nabla\cdot\left\{\left(\frac{\rho u^2}2+\frac{\gamma p_i}{\gamma-1}+\frac{B^2}{\mu_0}\right){\bf
  u}-\frac{{\bf B}({\bf u}\cdot{\bf B})}{\mu_0}\right\}=\nonumber\\
= -({\bf u}\cdot\nabla)\left(p_e+p_\sA\right)+
\frac{N_eN_ik_B}{\gamma-1}\left(\frac{\nu_{ei}}{N_i}\right)\left(T_e-T_i\right)-\frac{GM_\sSun\rho\,{\bf r}\cdot{\bf u}}{r^3}+Q_i,
\end{eqnarray}
\begin{eqnarray}\label{eq:electron}
\frac{\partial}{\partial
  t}\left(\frac{p_e}{\gamma-1}\right)&+&\nabla\cdot\left(\frac{p_e}{\gamma-1}{\bf
  u}\right)+p_e\nabla\cdot{\bf u}=\nonumber\\
&=&-\nabla\cdot\mathbf{q}_e
+\frac{N_eN_ik_B}{\gamma-1}\left(\frac{\nu_{ei}}{N_i}\right)\left(T_i-T_e\right)-Q_{\rm rad}+Q_e \,,
\end{eqnarray}
where $T_{e,i}$ are the electron and ion temperatures, $N_{e,i}$ are the electron and ion number densities, and $k_B$ is the Boltzmann constant. Other newly introduced terms are explained below.

The ideal equation of state, $p_{e,i}=N_{e,i}k_B T_{e,i}$, is used for both species. The polytropic index is $\gamma = 5/3$. The optically thin radiative energy loss rate in the lower corona is given by
\begin{equation}
Q_{\rm rad}=N_eN_i\Lambda(T_e) \,,
\end{equation}
where
$\Lambda(T_e)$ is the radiative cooling curve taken from the CHIANTI v7.1 database \citep[and references therein]{Landi:2013a}. The energy exchange rate between ions and electrons due to Coulomb collisions is defined in terms of the collision frequency
\begin{equation}
\frac{\nu_{ei}}{N_i}= \frac{2\sqrt{m_e} \Lambda_C
        (e^2/\varepsilon_0)^2}{3 m_p(2\pi k_BT_e)^{3/2}} \,,
\end{equation}
where $m_e$ and $e$ are the electron mass and charge, $m_p$ is the proton mass, $\varepsilon_0$ is the vacuum permittivity, $\mathbf{b}=\mathbf{B}/B$, and $\Lambda_C$ is the Coulomb logarithm. Finally, the electron heat flux $\mathbf{q}_e$ is expressed in the collisional formulation of \citet{Spitzer:1953a}:
\begin{equation}
\mathbf{q}_e = \kappa_\|\mathbf{bb}\cdot\nabla T_e,\quad
\kappa_\|=3.2\frac{6\pi}{\Lambda_C}\sqrt{\frac{2\pi}{m_e}\frac{\varepsilon_0}{e^2}^2}
\left(k_BT_e\right)^{5/2}k_B \,.
\end{equation}

\subsection{Transport and dissipation of \alf wave turbulence}
\label{sec:igor5}

Describing the dynamics of \alf wave turbulence and its interaction with  the background plasma requires special consideration.  The evolution of the \alf wave amplitude (velocity, $\delta\mathbf{u}$, and magnetic field, $\delta\mathbf{B}$) is usually treated in terms of the \citet{Elsasser:1950a} variables, $\mathbf{z}_\pm = \delta \mathbf{u}\mp {\delta\mathbf{B}}/{\sqrt{\mu_0\rho}}$. The Wentzel--Kramers--Brillouin (WKB) approximation \citep{Wentzel:1926a, Kramers:1926a, Brillouin:1926a} is used to derive the equations that govern transport of \alf waves, which may be reformulated in terms of the wave energy densities, $w_\pm=\rho\mathbf{z}^2_\pm/4$. Dissipation of \alf waves, $\Gamma_\pm w_\pm$, is the physical process that drives the solar wind and heats the coronal plasma.

\alf wave dissipation occurs when two counter-propagating waves interact. \alf wave reflection from steep density gradients is the physical process that results in local wave reflection, thus maintaining a source of both types of waves.
In order to describe this wave reflection we need to go beyond the WKB approximation that assumes that the wavelength is much smaller than spatial scales of the background variations.

The equation describing the propagation, dissipation, and reflection of \alf turbulence has been derived in \citet{vanderholst:2014a}:
\begin{equation}
\label{eq:w_pm}
\frac{\partial w_\pm}{\partial t} + \nabla\cdot\left[
(\mathbf{u}\pm\mathbf{V}_A)w_\pm
\right]
+\frac{w_\pm}{2}\left(\nabla\cdot\mathbf{u}\right) =
-\Gamma_\pm w_\pm\mp\mathcal{R}\sqrt{w_-w_+} \,,
\end{equation}
where $\mathbf{V}_A = \mathbf{B}/\sqrt{\mu_0 \rho}$ is the \alf velocity, while the dissipation rate ($\Gamma_\pm$) and the reflection coefficient (${\cal R}$) are given by
\begin{equation}
\Gamma_\pm = \frac{2}{L_\perp}\sqrt{\frac{w_\mp}{\rho}}
\end{equation}
and 
\bea
{\cal R}=\min\left\{\sqrt{\left({\bf b}\cdot[\nabla\times{\bf
    u}]\right)^2+\left[({\bf V}_A\cdot\nabla)\log
    V_A\right]^2},\max(\Gamma_\pm)\right\}\times\nonumber\\
\times\left[\max\left(1-\frac{I_{\rm max}}{\sqrt{{w_+}/{w_-}}},0\right)-\max\left(1-\frac{I_{\rm max}}{\sqrt{{w_-}/{w_+}}},0\right)\right] \,.
\eea
Here $L_\perp$ is the transverse correlation length of \alf waves in the plane perpendicular to the magnetic field line and $I_{\rm max}=2$  is the maximum degree of turbulence ``imbalance.'' If $\sqrt{w_\pm/w_\mp}<I_{\rm max}$, then \alf wave reflection is neglected and ${\cal R}=0$.

With the help of  the dissipation rate of \alf turbulence one can express the ion and electron heating rates:
\begin{equation}
Q_i = f_p \left(\Gamma_-w_- +\Gamma_+w_+\right),\quad
Q_e = (1-f_p)\left(\Gamma_-w_- +\Gamma_+w_+\right) \,,
\end{equation}
where $f_p\approx0.6$ is the fraction of \alf wave energy dissipated to the ions.

Finally, to close the system of equations, we use the following boundary condition for the Poynting flux of \alf waves, $\mathit{\Pi}_\sA$:
\begin{equation}
\frac{\mathit{\Pi}_\sA}{B} =\frac{\mathit{\Pi}_\sA({R_\sSun})}{B({R_\sSun})} = {\rm const }\approx
1.1\times10^6 \quad \left[\frac{\rm W}{{\rm m}^2{\rm T}}\right] \,.
\end{equation}
The transverse correlation length is assumed to scale with the magnetic field magnitude \cite[\eg][]{Hollweg:1986a}:
\begin{equation}
L_\perp\sim B^{-1/2},\quad
100 \ \left[{\rm km}\cdot{\rm T}^{1/2}\right] \leq L_\perp\sqrt{B} \leq 
300\ \left[{\rm km}\cdot{\rm T}^{1/2}\right] \,.
\end{equation}

\subsection{Modeling the transition region}
\label{sec:igor6}

\subsubsection{Chromosphere boundary conditions}

A simulation model based on the \alf wave turbulence may be extrapolated down to the top of the chromosphere. In order to save computational resources for this physics-based  model (which would require a gigantic amount of resources), the temperature and density at the top of the chromosphere are specified as:
\begin{equation}\label{eq:trhobc}
T_{ch}=(2\approx5)\times 10^4\,{\rm K},\qquad N_{ch}\approx2\times
10^{16}\,{\rm m^{-3}}.
\end{equation}
One needs to use an innovative approach to handle the sharp density gradients that have a spatial scale length of 
\begin{equation}\label{eq:chromoL}
L=\frac{k_BT_{ch}}{m_ig}\approx T_{ch}\times(30\ {\rm m/K}),
\end{equation}
which would greatly complicate the \alf wave turbulence model and introduce unmanageable wave reflection. To avoid this problem, we  apply WKB \alf wave turbulence effects and let the \alf waves freely propagate through the plasma at $T\le T_{ch}$. To both balance the radiative cooling and ensure the hydrostatic equilibrium, we apply an exponential heating function, $Q_h=A\exp(-x/L)$, to maintain the \textit{analytical} solution of the momentum and heat transfer equations, as follows:
$$
T_e=T_i=T_{ch},\qquad
N_e=N_i=N_{ch}
\exp\left(
-\frac{m_igx}{k_B(T_e+T_i)}\right),
$$
\begin{equation}
Q_h=Q_{rad}=N_e^2\Lambda(T_{ch})=N_{ch}^2\Lambda(T_{ch})
\exp\left(-\frac{m_igx}{k_BT_{ch}}\right).
\end{equation}
Here $g=274\ {\rm m/s^2}$ is the gravity acceleration near the solar surface,
the direction of this acceleration being antiparallel to the x-axis, and
$m_i$ is the proton mass. The two constants in the solution,
$N_{ch}$ and $T_{ch}$, which are the boundary values for the density and
temperature, respectively, are unambiguously related to the
amplitude, $A$,  of the heating function:
\begin{equation}
A=N_{ch}^2\Lambda(T_{ch}),
\end{equation}
and to the scale-length (see \equationname\ref{eq:chromoL}). Notice that there is a very simple relationship for the exponential scale-length for the heating function, which is half of the barometric scale-length of density variation: $2L=L_g=k_B(T_e+T_i)/(m_ig)$. 

The solution satisfies the equation for the heat conduction
as long as the heat transfer in the isothermic solution is absent and
heating at each point exactly balances the radiation cooling. The
hydrostatic equilibrium is also maintained, as long as
\begin{equation}
k_B\frac{\partial(N_eT_e+N_iT_i)}{\partial x}=-gN_im_i.
\end{equation}

The suggested solution does a good job describing the chromosphere. The short
scale-length of the heating function, (see \equationname\ref{eq:chromoL}), which is equal to $\approx0.6$ Mm
for $T_{ch}=2\times10^4$,  may presumably mimic absorption of
(magneto)acoustic turbulent waves, rapidly damping due to the
wave-breaking effects.  Physically, including this chromosphere heating 
function would imply that the temperature in the chromosphere is elevated 
compared to the photospheric temperatures due to some mechanism acting
in the chromosphere itself. By no means can this energy be transported
from the solar corona as long as the electron heat conduction rate at
chromospheric temperatures is very low.  

\subsubsection{Transition region}
\label{sec:TR}
One of the first successful models describing the Transition Region (TR), based on an \textit{analytical} solution, was published by
\cite{lion01} \citep[see also][]{lion09, Downs:2010a}. 
This model treats the TR as a thin continuous layer and it does not agree well with observations. This discrepancy suggests that the TR could be more accurately described as a carpet of 1D-like ``threads'' \citep[maybe spicules, see ][]{cran13}. Collectively these threads give the impression of a ``thin'' layer described by a solely height-dependent 1D solution.
To derive this solution one uses 1D governing equations to close the MHD model with the boundary condition at ``low boundary", which is at the same time the top boundary for the TR model. By solving the said 1D equations, one can merge the MHD model to the chromosphere, which is the bottom of the TR.

The heat transfer equation for a steady state hydrogen plasma in 
a uniform magnetic field reads:
\begin{equation}\label{transition}
\frac\partial{\partial s}\left(\kappa_0T_e^{5/2}\frac{\partial
  T_e}{\partial s}\right) +Q_h-N_e^2\Lambda(T_e)=0.
\end{equation}
Here $Q_h=\Gamma(w_-+w_+)$ is the coronal heating function, assumed to be constant at 
 spatial scales typical for the TR. 
Note that the coordinate is taken along the
magnetic field line, not along the radial direction.

By multiplying \equationname(\ref{transition}) by $\kappa_0T_e^{5/2}(\partial
T_e/\partial s)$, and by integrating from the interface to the chromosphere
 at temperature $T_e$, one can obtain:
\begin{equation}\label{transition1}
[
\frac12
\kappa_0^2 T_e^5
\left(
\frac{\partial T_e}{\partial s}
\right)^2
+\frac27\kappa_0 Q_h T_e^{7/2}
]|^{T_e}_{T_{ch}}=
(N_eT_e)^2
\int^{T_e}_{T_{ch}}{\kappa_0T^{1/2}\Lambda(T)dT}.
\end{equation}
Here the product, $N_eT_e$, is assumed to be constant. Therefore, it is
separated from the integrand. 
For a given $T_{ch}$, the only parameter in the solution is
$(N_eT_e)$. It can be expressed at any point in terms of the local
value of the heating flux and the radiation loss integral:
\begin{equation}\label{transition2}
(N_eT_e)=\sqrt{\frac{\frac12
\kappa_0^2 T_e^5
\left(
\frac{\partial T_e}{\partial s}
\right)^2
+\frac27\left(\kappa_0 Q_h T_e^{7/2}-\kappa_0 Q_hT_{ch}^{7/2}\right)}{\int^{T_e}_{T_{ch}}{\kappa_0T^{1/2}\Lambda(T)dT}}}.
\end{equation} 

The {\it  assumption} of constant $(N_eT_e)$ is fulfilled only if the effect
of gravity is negligible. Quantitatively, this condition is not 
trivial, as long as both the barometric scale and especially the heat
conduction scale are  functions of temperature. The barometric scale
may be approximated as $L_g(T_e)\approx T_e\times(60\ {\rm m/K}) $. The heat conduction scale, $L_h$, can be
estimated by noticing that within a large part of the transition
region the radiation losses dominate over the heating function,
so they are balanced by heat conduction: $\kappa_0
T_e^{5/2}\times(Te/L_h^2)\sim Q_r$. Thus, the condition for
neglecting gravity is:
\begin{equation}\label{restriction}
L_g(T_e)\approx T_e\cdot(60\ 
{\rm m/K})\gg 
L_h\approx
\sqrt{ \frac{\kappa_0T_e^{9/2}}{\Lambda(T_e)(N_eT_e)^2}}.
\end{equation}

\begin{figure}
\begin{center}
\includegraphics[width=0.6\textwidth]{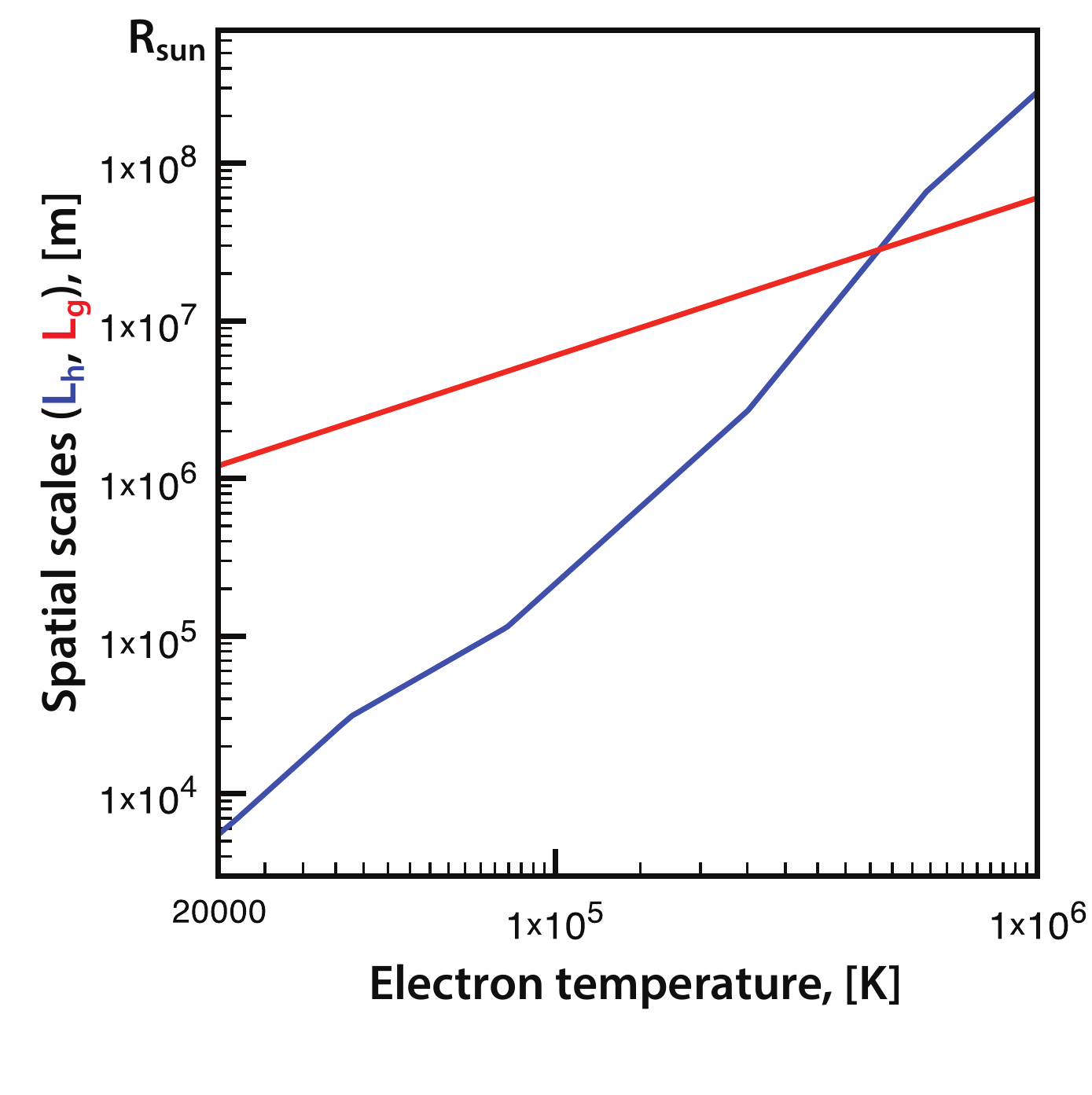}
\caption{Typical scales of the transition region: the heat
conduction scale (blue), $L_h$, and the gravitational height (red), $L_g$.}
\label{fig:length}
\end{center}
\end{figure} 

In \figurename~\ref{fig:length} we plot temperature dependencies $L_h(T_e)$ and $L_g(T_e)$ for $(N_eT_e)=10^{20}\ {\rm K/m^3}$. We see that near the
chromosphere boundary the approximation given by \equationname(\ref{restriction}) works very well as long as the temperature changes with height are very abrupt. The increase in temperature to $10^5{\rm K}$ occurs in less than 0.1 Mm. This estimate agrees with the temperature profile seen in the chromospheric lines (see, e.g., Fig.~2 and Fig.~8 in \cite{avre08}).  However, as the temperature increases with height, the effect of gravity on the temperature and density profiles becomes more significant. It becomes comparable to the heat
conduction effect at $T_e\approx 4.5\times 10^5$ K, which can be accepted
as the coronal base temperature, so that  the transition region corresponds to the temperature range from $T_{ch}\approx 2\times 10^4$ K to $4.5\times10^5$ K, with a typical width of $\sim 10\ {\rm Mm}\approx R_S/70$. The TR solution merges to the chromosphere solution with no jump in pressure. The merging point in the chromosphere, therefore, is at the density of $(N_eT_e)/T_{ch}\sim 10^{16}\ {\rm cm^{-3}}$. The short heat conduction scale at the chromosphere temperature (see \figurename~\ref{fig:length}) ensures that the heat flux from the solar corona across the transition region does not penetrate to higher densities.

In Section \ref{sec:TFLM} we revisit the transition region analytical model. 
However, in Section \ref{sec:awsom} we use another way to model the transition region, by artificially increasing the heat conduction in the lower temperature range \cite[see][]{abbe07}.  
Consider the transformation of the temperature functions shown in  
Eqs.(\ref{transition}--\ref{transition1}):
\begin{equation}
\kappa_0\rightarrow f\kappa_0,\qquad ds\rightarrow f\ ds,\qquad
\Gamma\rightarrow\Gamma/f,\qquad Q_{rad}\rightarrow Q_{rad}/f,
\end{equation}
with a common factor, $f\ge1$. The equations do not change in this 
transformation and the only effect on the solution is that the
temperature profile in the transition region becomes a factor of $f$
wider. By applying the factor, $f=(T_m/T_e)^{5/2}$ at $T_{ch}\le T_e\le
T_m$, the heat conduction scale in this range is
almost constant and is close to $\approx 1$ Mm for a choice of
$T_m\approx 2.5\times 10^5$ K (see \figurename~\ref{fig:length}).

It should be emphasized, however, that the choice of temperature
range for this transformation is highly confined by the
condition given in \equationname(\ref{restriction}). If a higher value of
$T_m$ is chosen, the heat conduction scale at the chromospheric temperature
exceeds the barometric scale in the chromosphere, resulting in a
unphysical penetration of the coronal heat into the deeper
chromosphere. The global model of the solar corona, with this
unphysical energy sink, suffers from reduced values for the coronal 
temperature and produces a visible distortion in the EUV and X-ray
synthetic images. {\it Thus, in formulating the transition region model
we modify the heat conduction, the radiation loss rate, and the wave
dissipation rate, and the maximum temperature for this modification does
not exceed $T_m\approx 2.5\times 10^5$ K.}


\section{The \alf Wave Solar Model (AWSoM)}
\label{sec:awsom}

\subsection{\alf wave-driven MHD coronal models} 
\label{sec:chip1}

With the help of the mathematical formalism described in Sect.~\ref{sec:igor2}, \alf wave-driven self-consistent models of the solar atmosphere and corona can be developed. The potential role of  low-frequency \alf waves to provide heat and momentum to accelerate the solar wind has been already recognized in the first decade of \textit{in-situ} exploration of the interplanetary medium.  \alf waves have long been measured \textit{in situ} in the  solar wind \citep{Belcher:1971a}, and have more recently been remotely observed  in the solar corona \citep{DePontieu:2007a, Cranmer:2009a}, where their energy is  sufficient to heat and accelerate the solar wind.  The theoretical exploration of \alf waves was first suggested in early work by  \cite{Hollweg:1978b, Hollweg:1981a, Hollweg:1982a}. Based on this early work, theories were developed that describe the  evolution and transport of Alfv\'enic turbulence, \eg \cite{Zank:1996a, Matthaeus:1999a, Zank:2014a, Zank:2017a}.
To self-consistently describe the heating and acceleration of the solar wind
with Alfv\'enic turbulence, several extended magnetohydrodynamic (MHD) 
models have been developed.  One-dimensional models include those by \eg
\cite{Tu:1997a, Laitinen:2003a, Vainio:2003a, Suzuki:2006a, Cranmer:2010a, Adhikari:2016a},
while multi-dimensional models include  \eg
\cite{Usmanov:2000a, Suzuki:2005a, Cranmer:2009a, vanderholst:2010a, Lionello:2014a}.  

These models have many common features.  First, they employ low-frequency 
\alf waves, which are assumed to dissipate below the ion cyclotron 
frequency.  Wave amplitudes are typically prescribed at the inner boundaries 
to match observed wave motions in the low corona \citep{DePontieu:2007a}. Wave 
energy propagates at the \alf speed along the magnetic field  
and drives the corona in two ways: (i) wave pressure gradient provides a 
volumetric force that accelerates the solar wind, while (ii) wave dissipation heats
the plasma.  In a variety of implementations, these \alf wave-driven models have
been shown to self-consistently reproduce the fast/slow solar wind speed 
distribution \citep[\eg][]{Usmanov:2000a, vanderholst:2010a, vanderholst:2014a}. 

An alternative method of coronal heating was developed by \citep{Suzuki:2002a, Suzuki:2004a}. Here, the focus is on shock wave heating instead of turbulent dissipation. The assumption is that slow and fast magneto-acoustic waves are generated by small scale reconnection events. These wave steepen into shocks while propagating along the field lines into the corona to heat and accelerate the plasma. \cite{Suzuki:2004a} demonstrated that the dissipation of shock trains can satisfactory reproduce the fast and slow wind speeds, except for the observed high temperatures in the slow wind, where other heating mechanisms might be needed. To the best of our knowledge, there are no 3D simulations that self-consistently include shocks and turbulence to assess which mechanism dominates in the heating and acceleration.

The recently developed \alf Wave Solar Model (AWSoM) 
\citep{Sokolov:2013a, vanderholst:2014a, Meng:2015a} extends the description  
with a three-dimensional solar corona/solar wind model that self-consistently
incorporates low-frequency \alf wave turbulence.  The model employs a 
phenomenological treatment of wave dissipation, with a prescribed correlation 
length inversely proportional to the magnetic field strength.  In this case, the wave spectrum is not resolved, 
so only the total forward and backward propagating wave energy densities and 
the partitioning of dissipated wave energy  between electrons and protons is 
fixed. Turbulence parameters are the wave energy densities, the correlation 
length, and the reflection rate.  The wave reflection model used is essentially
the same formulation as introduced by \cite{Matthaeus:1999a}.  In this case, 
the energy of the dominant wave is transferred to a counter-propagating minor 
wave, with the reflection coefficient controlled by the gradient of the 
\alf speed.

With AWSoM, \alf waves are represented as two discrete
populations propagating parallel and antiparallel to the magnetic field,
which are imposed at the inner boundary with a Poynting flux of the 
outbound \alf waves assumed to be proportional to the magnetic field 
strength.  The wave spectrum is not resolved, so the waves are presented
as frequency-integrated wave energies that propagate parallel to the
magnetic field at the local \alf speed.  The waves possess a pressure
that does work and drives the expansion of the plasma.  In this model,
outward propagating waves experience partial reflection on field-aligned
\alf speed gradients and the vorticity of the background.
The partial reflection leads to nonlinear interaction between oppositely
propagating \alf waves and results in an energy cascade from the 
large outer scale through the inertial range to the smaller perpendicular 
gyroradius scales, where the dissipation takes place. The apportioning 
of the dissipated wave energy to the isotropic electron temperature and 
the parallel and perpendicular proton temperatures depends on criteria 
for the particular kinetic instabilities that are involved \citep{Meng:2015a}. 
To apportion heating to the various ion species, we use the multispecies 
generalization of the stochastic heating, as described by \cite{Chandran:2013a}.

In the AWSoM model, the partitioning strategy is based on the dissipation 
of kinetic \alf waves (KAWs) with the stochastic heating mechanism 
for the perpendicular proton temperature \citep{Chandran:2011a}. 
In this mechanism, the electric field fluctuations due to
perpendicular turbulent cascade can disturb the proton gyro motion
enough to give rise to perpendicular stochastic heating, assuming that
the velocity perturbation at the proton gyroradius scale is large
enough. The firehose, mirror, and ion-cyclotron instabilities, due to
the developing proton temperature anisotropy, are accounted for. When
the plasma is unstable because of these instabilities, the parallel and
perpendicular temperatures are relaxed back toward marginal stable
temperatures, with the relaxation time inversely proportional to the
growth rate of these instabilities. See the work of \cite{Meng:2012a, Meng:2015a} for 
a detailed description. In this global model, excess of energy in the lower  
corona is transported back to the upper chromosphere via electron heat 
conduction where it is lost via radiative cooling.  

The AWSoM model is representative of the state of the art of extended MHD 
\alf wave-driven coronal models,  presenting many significant advances 
in modeling capability. 
First, turbulent dissipation rates are based directly on counter propagating
wave amplitudes, which are greatly enhanced by wave reflection at Alfv\'en
speed gradients.  Second, the model captures temperature anisotropies caused 
by preferential perpendicular heating in the fast solar wind.  Third, 
the effects of kinetic instabilities: fire hose, mirror mode, and cyclotron
instabilities limit temperature anisotropies with thresholds that are dependent 
on the proton temperature ratio and plasma $\beta$. Finally the three-dimensional 
model includes the entire structure of the corona including active regions and
slow and fast streams. This is the first time such kinetic physics has been 
incorporated into a global numerical model of a CME propagating through the 
solar corona, which allows us to address both particle heating, \alf wave 
damping, and their nonlinear coupled interaction as shown in \cite{Manchester:2017b}. 

\subsection{Multi-temperature coronal models}

Magnetohydrodynamic (MHD) theory is the simplest self-consistent model 
describing the macroscopic structure of the corona comprising the global 
distribution and temperature of the coronal plasma and magnetic field.
Such MHD models ignore the extreme complexity of a coronal environment that 
is made up of many plasma species affected by a wide range of wave-particle 
and particle-particle interactions, where heating occurs by the dissipation
of waves and time varying electric currents.  Particle populations are far 
from equilibrium and exhibit vastly different temperatures and distribution 
functions with extended high energy tails, the full complexity of which can 
only be described by kinetic models with non-Maxwellian velocity distribution 
functions \citep{Landi:2003a}.  For electrons, there are two nearly isotropic 
populations: the thermal core and the suprathermal halo, and a field aligned
strahl component \citep{Rosenbauer:1977a} that travels away from the Sun.  
Ions are more often characterized by a population that is anisotropic with 
a temperature perpendicular to the magnetic field higher than that parallel 
to the field.  Hydrogen is fully ionized, and all other atomic species are highly 
ionized. Protons, being almost 2000 times more massive than electrons, 
thermodynamically decouple at a distance of roughly 1.5 solar radii where 
Coulomb collisions become infrequent.  
 
To begin to address a range of physical processes as well as reduce the 
number of free parameters and ad hoc assumptions, a new generation of 
extended MHD global coronal models were developed.  First and foremost, 
thermodynamic processes were added, beginning with heat conduction 
and radiative losses, which allowed models to accurately capture 
the temperature structure of the lower atmosphere.  The use of stretched 
radial grids allow these models to resolve the transition region so 
that they may extend down to the upper chromosphere 
\citep{Lionello:2009a, Downs:2010a, Sokolov:2013a, vanderholst:2014a}. 
The radiative looses for these models are almost universally based on
CHIANTI tables \citep[]{Dere:1997a}, which specify the optically thin 
losses from the corona, which is dominated by line emission from heavy 
ions impacted by thermal electrons.  

With the thermal processes captured, extended MHD simulations can 
successfully reproduce images of the low corona provided by extreme
ultraviolet imaging telescopes, including SOHO/EIT, STEREO/EUVI, and SDO/AIA.
Observations provided by \cite{Downs:2010a} and 
\cite{vanderholst:2014a} indicate the qualitative match to coronal temperature and density available with the new models.  Figure~\ref{fig:EUV} 
provides an example of coronal ultraviolet images from \cite{Jin:2017a}. Here, 
the simulated active regions for 7 March 2011 (CR2107) naturally produce the enhanced emissions observed by
SDO/AIA 211{\AA}, STEREO A/EUVI 171{\AA}, and STEREO B/EUVI
195{\AA}.
The distinct feature in the present model is the enhanced wave reflection in
the presence of strong magnetic fields, such as in close proximity to active
regions that can increase the dissipation and thereby intensify the
observable EUV emission.

\begin{figure}[htbp]
\centering
\includegraphics[width=0.8\textwidth]{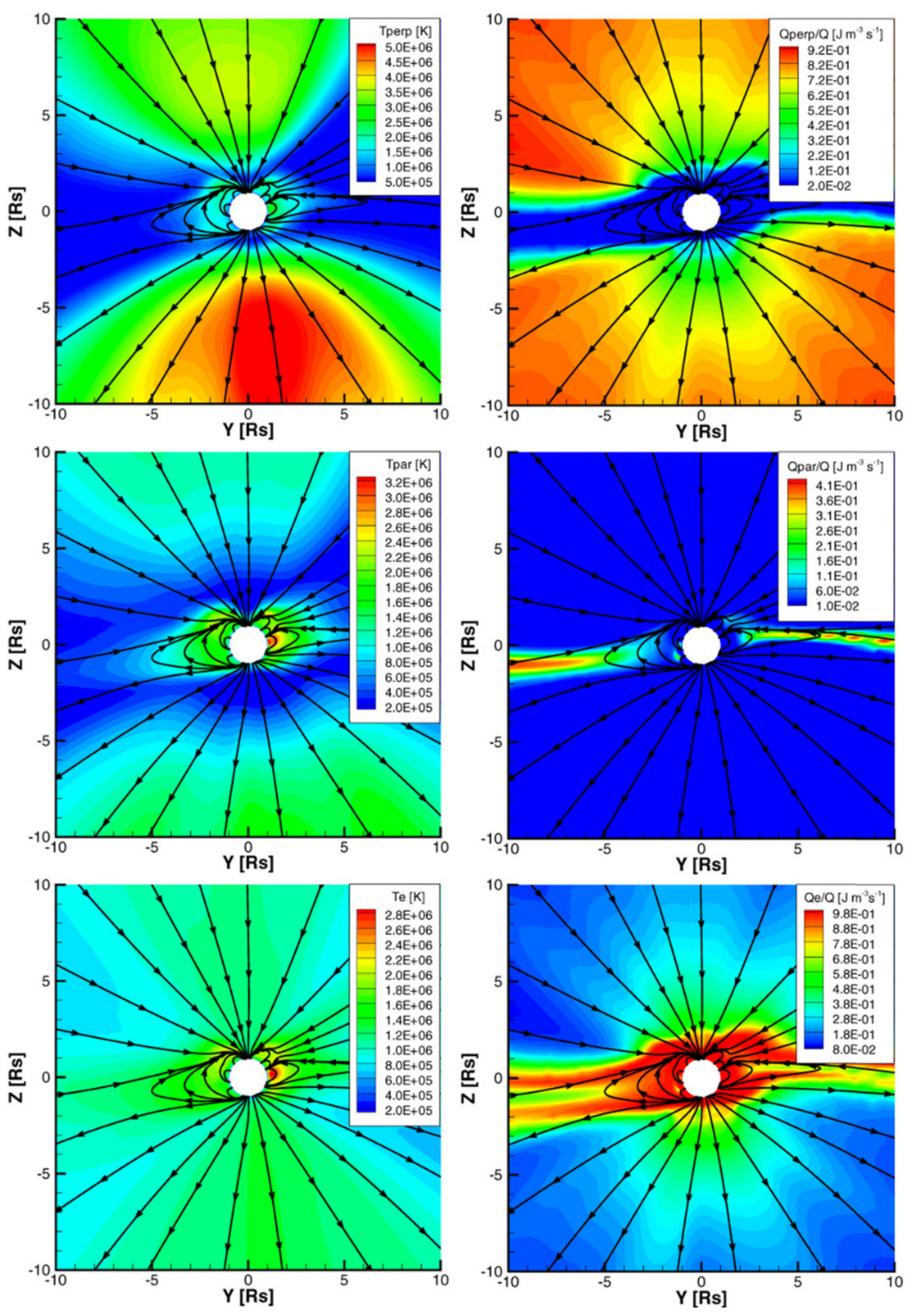}
\caption{Temperature and heating rates for the three-temperature steady-state solar wind solution adapted from \cite{vanderholst:2014a}.  Left panels show (top to bottom, respectively) color images of perpendicular and parallel proton temperatures and electron temperatures. Magnetic field lines are shown, ignoring the out-of-plane component. Right panels show (top to bottom, respectively) the fractions of perpendicular and parallel proton heating and electron heating by turbulent dissipation.}
\label{fig:3Twind}
\end{figure}


Even more complex, nonequilibrium thermodynamics can be captured with multiple-temperature single-fluid coronal models.  Two-temperature coronal models describe protons and electrons with a single fluid 
velocity but with individual energy equations and temperatures 
\cite[\eg][]{Sturrock:1966a, vanderholst:2010a}. 
The impetus for the feature stems from two facts: First, protons are almost 2000 times more massive than electrons, so that at one million degrees, their respective sound speeds are 120 km/s and 5000 km/s. Second, Coulomb collisions are so infrequent that within a fraction of a solar radius above the surface the ions and electrons thermally 
decouple from each other.  Consequently, heat conduction in the corona is completely dominated by electrons, which is particularly conspicuous in CME-driven shocks.  

The speed of fast CMEs occurs in the range where protons are shocked but not the electrons, and beyond a distance of two solar radii ($R_{\sSun}$) collisions become so infrequent that protons and electrons thermally decouple on the timescale of the shock 
passage.  As a result, protons can be shock heated to high temperature, while in the same location electrons cool from adiabatic expansion and heat conduction.  These temperature structures in CMEs were first modeled with one-dimensional two-temperature simulations by \cite{Kosovichev:1991a} and \cite{Stepanova:2000a}, and later in three-dimensional simulations by \cite{Manchester:2012a} and \cite{Jin:2013a}.  

The three-temperature thermodynamics model captures the electron temperature and resolves proton temperature into components parallel and perpendicular to the magnetic field.  Such models can capture the temperature anisotropy produced by a nearly collisionless plasma heated by wave-particle  interaction.  A leading example is the three-temperature version of AWSoM described in \cite{vanderholst:2014a} and \cite{Meng:2015a} and shown in \figurename~\ref{fig:3Twind}.  Here, particle temperatures and heating rates are shown on the left and right, respectively, and are determined by the incorporation of a description of turbulent dissipation developed by \cite{Chandran:2013a}.  This theory describes the turbulent cascade and dissipation of kinetic \alf waves, providing the thermal energy partitioning between protons and electrons.  As seen in \figurename~\ref{fig:3Twind}, for regions of low plasma beta, such as coronal holes, most energy goes to perpendicular proton heating, while in high beta regions, such as the current sheet, parallel heating dominates.  Electron heating dominates at intermediate beta levels found at the margins of the current sheet.  In AWSoM, temperature anisotropies are limited by kinetic instabilities, which are invoked when temperature ratios surpass the instability thresholds of fire hose, mirror mode, and cyclotron kinetic instabilities.  This three-temperature model has also been applied to study the  thermodynamics and the interaction \alf turbulence with CMEs and CME-driven shocks \citep{Manchester:2017b}.


\section{Threaded field line model}
\label{sec:TFLM}

\subsection{Magnetic ``treads''}
\label{sec:thread}

In the transition region the plasma temperature increases some two orders of magnitude over $\sim10^2$ km, resulting in a temperature gradient of $\sim10^4$ K/km. To resolve this gradient 3-D numerical simulations require sub-kilometer grid spacing, making these simulations computationally very expensive.

An alternative approach is to reformulate the mathematical problem in the region between the chromosphere and the corona in a way that decreases the computational cost. Instead of solving a computationally expensive 3-D problem on a very fine grid, one can reformulate it in terms of a multitude of much simpler 1-D problems along  \textit{threads} that allows 
us to map the boundary conditions from the the solar surface to the corona.
This approach is called the \textit{Threaded-Field-Line Model} (TFLM) \citep{Sokolov:2016a}.

The physics behind the reformulated problem is the assumption that between the solar surface and the top of the transition region ($R_\sSun<r<R_{b}$) the magnetic field can be described with a scalar potential. A \textit{thread} represents a field line of this potential field. One can introduce a 1-D problem that describes a magnetic flux tube around a given thread \citet{Sokolov:2016a}.

The magnetic field is divergenceless, therefore the magnetic flux remains constant along each thread:
\begin{equation}
  \label{eq:awsomr:flux}
  B(s) \cdot A(s)={\rm const} \,,
\end{equation}
where $s$ is the distance along the field line, and $A(s)$ is the cross-section area of the flux tube. Other conservation laws are also greatly simplified due to the fact that in a low-beta plasma, the flow velocity is aligned with the magnetic field. Assuming steady-state, the basic conservation laws can be written as 1-D equations.\\
Continuity equation:
\begin{equation}
\frac{\partial}{\partial s}\left(\frac{\rho u}{B}\right)=0\quad\Rightarrow\quad
\left(\frac{\rho u}{B}\right)={\rm const} \,.
\end{equation}
Conservation of momentum:
\begin{equation}
\frac{\partial p}{\partial s} = - \frac{b_r G M_\sSun \rho}{r^2} 
   \quad \Rightarrow \quad
p=p_ {\scriptscriptstyle{\rm TR}} \exp \left[{\int\limits_{R_ {\scriptscriptstyle{\rm TR}}}^r
\frac{GM_\sSun m_p}{2k_B T(r^\prime)}d\left(\frac{1}{r^\prime}\right)}\right],
\end{equation}
here $p=p_i+p_e$, $2T = T_i+T_e$, $R_ {\scriptscriptstyle{\rm TR}}$ is the radius of the bottom of the transition region (TR), and $b_r$ is the radial component of $\mathbf{b}$. In this expression terms proportional to $u^2$ are neglected, $\mathbf{j}\times\mathbf{B}$ is omitted due to the fact that electric currents vanish in a potential field ($\mathbf{j} \propto \nabla \times \mathbf{B} = 0$), and the pressure of \alf wave turbulence is assumed to be much smaller than the thermal pressure, $p_\sA\ll p$.\\
Conservation of energy:
\bea
\lefteqn{
\frac{2N_ik_B}{B\left(\gamma-1\right)}\frac{\partial T}{\partial t} + 
\frac{2k_B\gamma}{\gamma-1}\left(\frac{N_iu}{B}\right)
\frac{\partial T}{\partial s} =
}\nonumber\\&&
\frac{\partial}{\partial s}\left(\frac{\kappa_\|}{B}
\frac{\partial T}{\partial s}\right) + 
\frac{\Gamma_-w_-+\Gamma_+w_+-N_eN_i\Lambda(T)}{B}+
\left(\frac{\rho u}{B}\right)
\frac{\partial}{\partial s}  \left(\frac{GM_\sSun}{r} \right) \,,
\eea
where the term ${\partial T}/{\partial t}$ is retained because it is assumed that the electron heat conduction is a relatively slow process.

In addition to the plasma equations, the \alf wave dynamics can also be reformulated.
In \equationname(\ref{eq:w_pm}), we introduce a new variable, $a_\pm^2$:
\begin{equation}
a_\pm^2 = \frac{V_A}{\mathit{\Pi}}w_\pm \,.
\end{equation}
With the help of this substitution, the \alf wave transport equation becomes
\begin{equation}
\frac{\partial a_\pm^2}{\partial t}+\nabla\cdot\left(\mathbf{u}a_\pm^2\right)
\pm\left(\mathbf{V}_A\cdot\nabla\right)a_\pm^2=
\mp\mathcal{R}a_-a_+
-  2 \sqrt{\frac{\mathit{\Pi}}{B}\frac{\mu_0 V_A}{\left(L_\perp\sqrt{B}\right)^2}}a_\pm^2 a_\mp \,.
\end{equation}
These equations can be additionally simplified since in the lower corona $u\ll V_A$
(i.e., waves are assumed to travel fast and quickly converge to equilibrium), therefore we can neglect the $\partial a_\pm^2/\partial t$ terms:
\begin{equation}
\pm\left(\mathbf{b}\cdot\nabla\right)a_\pm^2=
\mp\frac{\mathcal{R}}{V_A}a_-a_+
-  2 \sqrt{\frac{\mathit{\Pi}}{B}\frac{\mu_0 V_A}{\left(L_\perp\sqrt{B}\right)^2}}a_\pm^2 a_\mp \,.
\end{equation}
Additionally, we introduce a new variable: 
\begin{equation}
d\xi = ds \sqrt{\frac{\mathit{\Pi}}{B}\frac{\mu_0}{V_A \left(L_\perp\sqrt{B}\right)^2}} \,.
\end{equation}
Now the wave equations become
\begin{equation}
\pm\frac{da_\pm}{d\xi} = \mp \frac{ds}{d\xi} \frac{\mathcal{R}}{2V_A} a_\mp - a_{-} a_{+} \,.
\label{eq:a-pm}
\end{equation}

Equations~(\ref{eq:a-pm}) describe boundary value problems and one needs to specify boundary conditions somewhere along the thread. Let $\xi_{-}$ denote the location of the lower boundary at the outgoing end of the thread (where the field direction points away from the Sun), and $\xi_{+}$ denote the lower boundary at the downward end of the thread. The boundary conditions now must specify the values of $a_\pm$ at the location where the \alf turbulence enters into the thread: $a_{+}(\xi=\xi_{-}) = a_{+_0}$ and $a_{-}(\xi=\xi_{+}) = a_{-_0}$. The values of $a_{\pm_0}$ are empirically specified.

\subsection{From the transition region to the threaded field line corona}
\label{sec:igor7}

Finally, one must specify the plasma and turbulence conditions at the interface between the threaded field line region  and the corona at the radial distance of $r=R_b$. These conditions depend on the direction of the magnetic field at the interface.\\

If the magnetic field points outward, $b_r(r=R_b)>0$:
\bea
\left(\frac{u}{B}\right)_{TF}=\quad
\left(\frac{\mathbf{u}\cdot\mathbf{B}}{B^2}\right)_{cor};
\quad\left(a_-\right)_{TF}= \left(a_-\right)_{cor};\quad
\left(a_+\right)_{cor}= \left(a_+\right)_{TF}\ \,.
\eea

If the magnetic field points inward, $b_r(r=R_b)<0$:
\bea
\left(\frac{u}{B}\right)_{TF}=-
\left(\frac{\mathbf{u}\cdot\mathbf{B}}{B^2}\right)_{cor};
\quad\left(a_+\right)_{TF}= \left(a_+\right)_{cor};\quad
\left(a_-\right)_{cor}= \left(a_-\right)_{TF} \,.
\eea

In addition, the temperature and density need to be matched at the interface between the threaded field line and the corona. In order to achieve this it is assumed that at the interface the temperature gradient is mainly in the radial direction:
\begin{equation}
\left(\frac{\partial T}{\partial r}\right)_{cor}=
\frac{1}{b_r} \left(\frac{\partial T}{\partial s}\right)_{TF} \,.
\end{equation}
The boundary condition for the density is controlled by the sign of $\mathbf{b}\cdot\mathbf{u}$:
\bea
{\rm for}\quad \mathbf{b}\cdot\mathbf{u}>0&:&\quad\left(\frac{N_iu}{B}\right)_{TF}=
\left(N_i\right)_{TF}\left(\frac{u}{B}\right)_{cor};
\nonumber\\
{\rm for}\quad \mathbf{b}\cdot\mathbf{u}<0&:&\quad\left(\frac{N_iu}{B}\right)_{TF}=
\left(\frac{N_iu}{B}\right)_{cor}.
\eea

In the last step, one needs to consider the energy balance in the transition region where two physical processes balance each other: heat conduction and radiative cooling. Assuming steady-state conditions, this energy balance can be expressed using \equationname(\ref{transition}) for $Q_h=0$,
\begin{equation}
\frac\partial{\partial s}\left(\kappa_0T^{5/2}\frac{\partial T}{\partial s} \right) =N_eN_i\Lambda(T) \,,
\end{equation}
where for the field-aligned heat conduction coefficient the usual $\kappa_\|=\kappa_0T^{5/2}$ expression is used.
The length of a given magnetic field line between the photosphere and the top of the transition region is obtained by integrating the thread:
\begin{equation}
L_{TR}=\int\limits_{R_\sSun}^{R_{TR}}ds \,.
\end{equation}
If the temperature at the top of the transition region, $T_{TR}$, is known, one can obtain the heat flux and pressure from the following equations:
\bea
N_i k_B T = \frac1{L_{TR}} \int_{T_{ch}}^{T_{TR}} {\frac{\kappa_0\tilde{T}^{5/2}d\tilde{T}}
{U_{\rm heat}(\tilde{T}) }} \,,
\eea
where $T_{ch}\approx(1\div2)\times10^4K$ and
\bea
\kappa_0 T_{TR}^{5/2} \left(\frac{\partial T}{\partial s}\right)_{T=T_{TR}} = N_i k_B T U_{\rm heat}(T_{TR}) \,.
\eea
Here 
\bea
U_{\rm heat}(T)=\sqrt{\frac2{k_B^2}\int^{T}_{T_{ch}}{  
\kappa_0(T^\prime)^{1/2}\Lambda(T^\prime)dT^\prime}} \,.
\eea
The $\Lambda(T)$ and $U_{\rm heat}(T)$ functions can be easily tabulated  using the CHIANTI database \citep{Dere:1997a, Landi:2013a}.

\section{Summary}

This review gives an historical introduction to large-scale modeling of the near-steady solar corona and the solar wind. It focuses on the ``quiet'' corona when the global structure is time-independent in the frame of reference corotating with the Sun. We start with an extensive -- and critical -- review of the early concepts of the solar corona and the \cite{Biermann:1951a}-\cite{Chapman:1957a} puzzle that led to Parker's (\citeyear{Parker:1958a}) revolutionary -- and highly controversial -- idea of the continuous solar wind.

Following the evolution of the solar wind concept, we describe the first numerical models of the expanding solar corona \citep{Noble:1963a,Scarf:1965a}, and the emergence of the potential magnetic field model \citep{Schatten:1968a, Schatten:1969a} and source surface model \citep{Altschuler:1969a, Schatten:1969a}.

Since the 1990s, significant progress has been made in describing and modeling the heliosphere. MHD turned out to be surprisingly successful in describing the solar wind from tens to hundreds of $R_\sSun$ \cite[cf.][]{Pizzo:1991a, Odstrcil:1999a, Groth:1999b, Linde:1998a, Pogorelov:2013a, Opher:2003a}. However, modeling the transition from the dense, cold photosphere to the hot super-Alfv\'enic corona is still challenging. The first generation of coronal models used simplified energetics \cite[cf.][]{Usmanov:1993a, Mikic:1999a, Cohen:2007a} with considerable success. ``Thermodynamic'' models use a realistic adiabatic index but empirical heating functions to heat and accelerate the solar wind \cite[cf.][]{Groth:1999a, Lionello:2001a}. The advantage of the thermodynamic approach is their ability to describe shock related phenomena.

The latest generation of coronal models use Alfv\'en waves to heat and accelerate the solar wind \cite[cf.][]{Usmanov:2000a, vanderholst:2014a}. While this approach has the promise to explain the origin of both the fast and slow solar wind states, it is still at a relatively early state of development and much work remains to be done. We urge the reader to stay tuned.

Though model validation goes beyond the scope of this paper, one can envision investigations to determine the veracity of the models described in this review. As it was pointed out, one can compare emission properties in the EUV and soft x-ray with observed signatures in the low corona. In the extended corona, many of Alfv\'enic effects may be more pronounced, such as non-thermal velocities and signatures of wave dissipation such as temperature anisotropies. The Parker Solar Probe is ideally designed to resolve these questions with high cadence observations of electromagnetic waves and particle distribution functions. One can also validate models by comparing their predictions of solar wind parameters with those from the WSA model.  In addition, efforts are now underway to develop an integrated model describing the acceleration and transport of solar energetic particles (SEPs) directly coupled with an Alfw\'en wind turbulence based solar wind model \cite[see][]{boro18}. Such a coupled model addresses the effects of wave turbulence on SEP transport. In particular, particle diffusion rates, and arrival times, are affected by the turbulence level in the solar corona.  Therefore, SEP observations may provide additional validation capability for the coronal model.


\begin{acknowledgements}
\label{sec:acknowledgements}

The work performed at the University of Michigan was partially supported by the National Science Foundation grants AGS-1322543 and PHY-1513379, NASA grant NNX13AG25G, the European Union's Horizon 2020 research and innovation program under grant agreement No 637302 PROGRESS. We would also like to acknowledge high-performance computing support from: (1) Yellowstone (ark:/85065/d7wd3xhc), provided by NCAR's Computational and Information Systems Laboratory, sponsored by the National Science Foundation, and (2) Pleiades, operated by NASA's Advanced Supercomputing Division.
\end{acknowledgements}


\bibliographystyle{spbasic}      
\bibliography{livingrev1,livingrev2}   

\end{document}